%% file: sea3.tex
\documentclass[a4paper,11pt]{article}   
\usepackage{authblk} 
\usepackage[top=2cm,bottom=2cm,left=2cm,right=2cm]{geometry}
\usepackage{multicol}
\usepackage{listings}
\usepackage{color}

\usepackage{xspace}
\usepackage{tikz}
\usepackage{caption}
\usepackage{subcaption}
\usepackage{graphicx}
\usetikzlibrary{mindmap}

\def\PROB{\mathbb{P}}
\def\EXP{\mathbb{E}}

\newcommand{\progname}{\textsf}
\newcommand{\compname}{\textsf} 

\newcommand{\HTCondor}{\progname{HTCondor}\xspace}
\newcommand{\MW}{\progname{MW}\xspace}
\newcommand{\lrs}{\progname{lrs}\xspace}
\newcommand{\prs}{\progname{prs}\xspace}
\newcommand{\tbb}{\progname{TBB}\xspace}
\newcommand{\bpp}{\progname{Bob++}\xspace}
\newcommand{\pth}{\progname{Pthread}\xspace}
\newcommand{\gcc}{\progname{C}\xspace}
\newcommand{\mpi}{\progname{MPI}\xspace}
\newcommand{\zram}{\progname{ZRAM}\xspace}
\newcommand{\minisat}{\progname{Minisat}\xspace}
\newcommand{\glucose}{\progname{Glucose}\xspace}
\newcommand{\glucosesyrup}{\progname{Glucose-Syrup}\xspace}
\newcommand{\plingeling}{\progname{plingeling}\xspace}
\newcommand{\lingeling}{\progname{lingeling}\xspace}
\newcommand{\treengeling}{\progname{treengeling}\xspace}

\newcommand{\mplrs}{\progname{mplrs}\xspace}
\newcommand{\mts}{\progname{mts}\xspace}
\newcommand{\mptopcom}{\progname{mptopcom}\xspace}
\newcommand{\topcom}{\progname{TOPCOM}\xspace}
\newcommand{\polymake}{\progname{polymake}\xspace}

\newcommand{\btree}{\progname{btree}\xspace}
\newcommand{\mtree}{\progname{mtree}\xspace}

\newcommand{\btop}{\progname{btopsorts}\xspace}
\newcommand{\mtop}{\progname{mtopsorts}\xspace}
\newcommand{\vr}{\progname{VR}\xspace}
\newcommand{\genle}{\progname{Genle}\xspace}
\newcommand{\gray}{\progname{grayspan}\xspace}
\newcommand{\graysp}{\progname{grayspspan}\xspace}
\newcommand{\mtsat}{\progname{mtsat}\xspace}
\newcommand{\mtsatglucose}{\progname{mtsat-glucose}\xspace}

\newcommand{\maic}{\compname{mai32}\xspace}

\newcommand{\mainew}{\compname{mai32abcd}\xspace}
\newcommand{\mainewer}{\compname{mai32ef}\xspace}

\newcommand{\search}{\progname{search}\xspace}

\newcommand{\Adj}{\textrm{Adj}}

\newcommand{\mydepth}{\ensuremath{\mathit{depth}}\xspace}
\newcommand{\mymaxdepth}{\ensuremath{\mathit{max\_depth}}\xspace}
\newcommand{\startvertex}{\ensuremath{\mathit{start\_vertex}}\xspace}
\newcommand{\maxnodes}{\ensuremath{\mathit{max\_nodes}}\xspace}
\newcommand{\nbudget}{\ensuremath{\mathit{node\_budget}}\xspace}
\newcommand{\sdata}{\ensuremath{\mathit{sdata}}\xspace}

\newcommand{\shared}{\ensuremath{\mathit{shared\_data}}\xspace}
\newcommand{\inputdata}{\ensuremath{\mathit{input\_data}}\xspace}
\newcommand{\myfalse}{\ensuremath{\mathbf{false}}\xspace}
\newcommand{\mytrue}{\ensuremath{\mathbf{true}}\xspace}
\newcommand{\mycount}{\ensuremath{\mathit{count}}\xspace}
\newcommand{\putoutput}{\ensuremath{\textrm{output}}\xspace}
\newcommand{\unexplored}{\ensuremath{\mathit{unexplored}}\xspace}
\newcommand{\lmin}{\ensuremath{\mathit{lmin}}\xspace}
\newcommand{\lmax}{\ensuremath{\mathit{lmax}}\xspace}
\newcommand{\myscale}{\ensuremath{\mathit{scale}}\xspace}

\newcommand{\numworkers}{\ensuremath{\mathit{num\_workers}}\xspace}
\newcommand{\mysize}{\ensuremath{\mathit{size}}\xspace}
\newcommand{\maxd}{\ensuremath{\mathit{maxd}}\xspace}
\newcommand{\mystart}{\ensuremath{\mathit{start}}\xspace}
\newcommand{\unfinished}{\ensuremath{\mathit{unfinished}}\xspace}

\newcommand{\pmtwotwo}{\ensuremath{\mathit{pm22}}\xspace}
\newcommand{\catfourtwo}{\ensuremath{\mathit{cat42}}\xspace}
\newcommand{\keightnine}{\ensuremath{K_{8,9}}\xspace}
\newcommand{\eightcage}{\mbox{\textit{8-cage}}\xspace}
\newcommand{\pfivecfive}{\ensuremath{P_5C_5}\xspace}
\newcommand{\cfivecfive}{\ensuremath{C_5C_5}\xspace}
\newcommand{\ksevenseven}{\ensuremath{K_{7,7}}\xspace}
\newcommand{\konetwo}{\ensuremath{K_{12}}\xspace}

\usepackage{todonotes}
\usepackage{amsmath}
\usepackage{amssymb}
\usepackage{latexsym}
\usepackage{graphicx}
\usepackage{color}
\usepackage{hyperref}
\usepackage{slashbox}
\usepackage{algorithm}
\usepackage{algpseudocode}

\definecolor{darkblue}{rgb}{0,0,0.6}


\include{./macros}
\begin{document}

\lstset{language=C, xleftmargin=\parindent, showspaces=false, showstringspaces=false} 

\title{\mts: a light framework for parallelizing tree search codes}
\author[1]{David Avis}
\author[2]{Charles Jordan}
\affil[1]{School of Informatics, Kyoto University, Kyoto, Japan and 
          School of Computer Science,
          McGill University, Montr{\'e}al, Qu{\'e}bec, Canada\\
          \texttt{avis@cs.mcgill.ca}}
\affil[2]{Graduate School of Information Science and Technology,
          Hokkaido University, Japan\\
          \texttt{skip@ist.hokudai.ac.jp}}

\maketitle

\begin{abstract}
We describe \mts, a generic framework for 
parallelizing certain types of tree search programs including reverse search,
backtracking, branch and bound and satisfiability testing.
It abstracts and generalizes the ideas used in parallelizing \lrs, a
reverse search code for vertex enumeration.
\mts supports sharing information between processes which is important for
applications such as satisfiability testing and branch-and-bound.
No parallelization is implemented in the legacy single processor programs
minimizing the changes needed and simplying debugging.  \mts is written
in \gcc, uses \mpi for parallelization and can be used on a network of computers.
We give four examples of reverse search codes parallelized by using \mts: topological
sorts, spanning trees, triangulations and Galton-Watson trees.
We also give a parallelization of two codes
for satisfiability testing. We give experimental results comparing
the parallel codes with other codes for the same problems.

\noindent{}Keywords: reverse search, parallel processing, topological sorts, spanning trees, 
triangulations, satisfiability testing\\
Mathematics Subject Classification (2000) 90C05
\end{abstract}

\section{Introduction}
\label{sec:intro}

Parallel programming is a vast area and there is a great amount
of literature on it (see, e.g., Mattson et al.~\cite{MSM}).
Topics include architecture, communication,
data sharing, interrupts, deadlocks, load balancing,
and the distinction between shared memory and distributed computing.
This is all essential for building an efficient parallel
algorithm from scratch. 

Our starting point was different.
We had a large complex code, \lrs, developed over
about 20 years and tested extensively, which solved vertex/facet enumeration problems.
These problems are notoriously hard and running times often take weeks or longer.
The underlying
algorithm, reverse search, was clearly suitable for parallelization.
Nevertheless, the mathematical intricacy of the underlying problem rendered the algorithmic
engineering of direct parallelization daunting. This led us to consider
building all of the parallelization into a wrapper, making only minor changes to the
underlying \lrs code. There followed a series of implementations resulting ultimately
in the authors' \mplrs code~\cite{AJ18a}. The key features of \mplrs are:
(a) there is no parallel code inside \lrs, (b) parallel threads execute \lrs on non-overlapping
subproblems, (c) there is no communication between threads except at the beginning
and end of a subproblem execution, (d) the computation can be distributed over a cluster
of computers, and (e) the wrapper is directly inserted into the \lrs library. 
Most of the topics in parallel computation mentioned
above are not major issues in this restricted framework. The exception is load balancing
for which we use a particularly simple method which consists of budgeting the number
of nodes evaluated in a subproblem.

It seemed likely that similar results could be obtained for other algorithms based on
reverse search\footnote{In 2008, John White made a list of 130 different applications and implementations, 
see link at \cite{tutorial}.} or similar easily parallelizable tree search methods. 
Many such sequential codes exist, so designing custom wrappers for each is not
desirable.
Our goal was to build a single generic 
wrapper that could be used, with little if any modification, to do the required 
parallelization while maintaining features (a)--(e) described above.
This resulted in \mts, presented here.  The current
implementation\footnote{Version used here available at \url{https://www-alg.ist.hokudai.ac.jp/~skip/mts/}}
uses MPI and works on clusters of machines.
The \mts framework is more general than \mplrs in that it allows the sharing of data 
obtained by subproblems, but still maintains the absence of communication between threads.
This widens its application to more general tree search problems such as satisfiablility testing
and branch and bound. 

In Section \ref{sec:survey} we survey the literature on parallelizing reverse search codes.
We then describe our general approach in Section \ref{sec:mts}
and apply it to reverse search in Section \ref{sec:app1}.
We give four examples:
generating topological sorts,
spanning trees of a graph, high dimensional triangulations and Galton-Watson trees. 
This latter problem was chosen as the trees generated are unbalanced, which is
a challenge for parallelization. They allow an analysis of the major source
of overhead in \mts and its dependence on the crucial budgeting parameter which will
be described in Section \ref{sec:mts}.

Tree search has wide uses, of which enumeration is just one
example. In fact it is a very specific example as all nodes in the enumeration tree
are visited.
Two other important uses of tree search are 
satisfiability testing and branch and bound.
Here the goal is {\em not} to search the entire tree but to prune subtrees when possible. The tree generated in these cases will normally
differ depending on the choices made at early stages and the sharing of information learned during the computation.
The \mts framework includes support for sharing data between processes and can be
applied to these types of problems. 
As an example we present a parallelization for satisfiability
testing in
Section \ref{sec:sat}, demonstrating how little of the original code needs to be changed. 

In Section \ref{sec:comp} we give computational results for parallelized 
reverse search and satisfiability codes.
This is followed in Section \ref{sec:hist} by a discussion of how to evaluate
the experimental results, the situation being quite different for enumeration problems and
for those problems where pruning is used. 
For the enumeration problems we get near linear speedup using several hundred cores.
For the satisfiability problem we show a substantial improvement in the number of SAT instances that
can be solved in a given fixed time period.
Finally we give some conclusions
and areas for future research
in Section \ref{sec:concl}. 

\section{Survey of previous work}
\label{sec:survey}

The reverse search method, initially developed for vertex enumeration, was extended to a wide
variety of enumeration problems \cite{AF93}.
From the outset it was realized that it was eminently suitable for parallelization.
In 1998, Marzetta announced his \zram parallelization
platform~\cite{BMFN99,ZRAMthesis} which can be used for reverse search, backtracking and branch and bound codes.
He successfully used it to parallelize several reverse search and branch and bound codes,
including \lrs from which he derived the \prs code.
Load balancing is performed using a variant of what is now known as job stealing.
Application codes, such as \lrs, were embedded into \zram itself leading
to problems of maintenance as the underlying codes evolved.                     
Although \prs is no longer distributed and was based on a now obsolete version of \lrs,
it clearly showed the potential for large speedups of reverse search algorithms. 

The reverse search framework in \zram was also used to implement a parallel code for
certain quadratic maximization problems~\cite{FKL05}.
In a separate project, 
Weibel \cite{Weibel10} developed a parallel reverse search code to compute Minkowski sums.
This C++ implementation runs on shared memory machines and he obtains linear speedups with up to 8 processors,
the largest number reported.

\zram is a general-purpose framework that is able to handle a number of other applications, such
as branch and bound and backtracking, for which there are by now a large number of competing frameworks.
This is a very large area and an extensive survey of those methods relevant to the present study
was given in \cite{AJ18a}. We give a brief overview here.
Recent papers by Crainic et al.~\cite{CLR06}, McCreesh et al.~\cite{MP15} and Herrera et al.~\cite{He17}
describe over a dozen such systems. While branch and bound may seem similar to reverse search enumeration, there
are fundamental differences. In enumeration it is required to explore the entire
tree whereas in branch and bound the goal is to explore as little of the tree as possible
until a desired node is found. The bounding step removes subtrees from consideration and
this step depends critically on what has already been discovered. Hence the order of
traversal is crucial and the number of nodes evaluated varies dramatically depending on this
order. Sharing of information is critical to the success of parallelization. 
These issues do not occur in reverse search enumeration, and so a much lighter wrapper is possible.

Relevant to the heaviness of the wrapper and amount of programming effort required,
a comparison of three frameworks is given in \cite{He17}. The first, \bpp  \cite{Dj06},
is a high level abstract framework, similar in nature to \zram, on top of which the application sits.
This framework provides parallelization with relatively little programming effort
on the application side and can run on a distributed network. 
The second, Threading Building Blocks (\tbb) \cite{Re07},
is a lower level interface providing more control but also 
considerably more programming effort. It runs on a shared memory machine. 
The third framework is the \pth model \cite{Ca08} in which parallelization
is deep in the application layer and migration of threads is done by the operating system.
It also runs on a shared memory machine.
All of these methods use job stealing for load balancing \cite{BL99}. 
In \cite{He17} these three approaches are applied to a global optimization algorithm.
They are compared on a rather small
setup of 16 processors, perhaps due to the shared memory limitation of the last two approaches. 
The authors found that \bpp achieved a disappointing speedup of about 3 times, 
considerably slower than the other two approaches
which achieved near linear speedup. 

A more sophisticated framework for parallelizing application codes over large 
networks of computers is \MW that works with the 
distributed environment of 
\HTCondor\footnote{Available at \url{https://research.cs.wisc.edu/htcondor/mw/}}.
\MW is a set of C++ abstract base classes that allow parallelization of 
existing applications based on the master-worker paradigm~\cite{Goux01}. We employ the same
paradigm in \mts although our load balancing methods are different. \MW
has been used successfully to parallelize combinatorial optimization problems such
as the Quadratic Assignment Problem, see the \MW home page for references.
Although \MW could be used to parallelize reverse search algorithms, we are not aware
of any such applications.

\section{The \mts framework}
\label{sec:mts}

The goal of \mts is to parallelize existing 
tree search codes with minimal internal modification of these codes.
The tree search codes should satisfy certain conditions, specified below.
The \mts implementation starts a user-specified number of
processes on a cluster of computers. One process becomes the \emph{master},
another becomes the \emph{consumer}, and the remaining are
\emph{workers} which essentially run the original tree search code
on specified subtrees. 
Communication is limited; workers are
not interrupted and do not communicate between themselves.

The master sends the input data and
parametrized subproblems to workers, informs the other processes to exit
when appropriate, and handles checkpointing.  The consumer
receives and synchronizes output.  Workers
get subproblems from the master, run the legacy
code, send output to the consumer, and return unfinished subproblems 
to the master.  

Generating subproblems can be done in many ways.
One way would be to report nodes at some initial fixed depth.
This works well for balanced trees
but many trees encountered in practice are highly unbalanced and the vast majority
of subtrees contain few nodes. Increasing the initial search depth does not
solve this problem. Ideally we would only break up the large subtrees and in the
development of \mplrs we tried various ways to estimate the size
of a given subtree. Experimentally this did not work well due to the high variance
of the estimator and the wasted cost of doing many estimates.

The idea that worked best, and is implemented in \mts, was also the simplest: a heuristic to determine
large subtrees called {\em budgeting}. 
This general approach is similar to but simpler than
the well-known work-stealing approach~\cite{BL99}.
When assigning work the master specifies
that a worker should terminate after completing a certain
amount of work, called a {\em budget},
and then return a list of unexplored subtrees.
The precise budget may depend on the application.
For enumeration problems it could be the number of nodes visited by the worker.
Some advantages of budgeting are:
\begin{itemize}
\item small subtrees are explored without being broken up
\item large subtrees will be broken up repeatedly
\item each worker returns periodically for reassignment,
can give information to be passed on to other workers and receive such information
\item it is implemented on-the-fly and avoids the duplication of work done in estimation
\item it can be varied dynamically during execution to control the job list size
\item when used statically and without pruning, 
      the overall job list produced is deterministic and independent of the number of workers
\end{itemize}
This last item is useful for debugging purposes and also enables a theoretical analysis of the job list size under
certain random tree models, see \cite{AD17a}.
In particular, methods that limit work based on time (such as ``begetting'' in \MW)
do not have this property. 

Implementing budgeting does not require interrupting workers or communication between workers.
The master uses dynamic budgets
to control the job list: small budgets break up more subtrees and lengthen
the joblist while large budgets have the reverse effect.

Additional features of \mts include checkpointing and restarts, allowing 
the user to move jobs or free computing
resources without losing work.
\mts can produce various histograms to help tune performance.
Histograms and their uses are described in
Section \ref{sec:hist}.  

\subsection{Sequential tree search code}
\label{subsec:seq}

To be suitable for parallelization with \mts the underlying
tree search code, which we will call \search, must satisfy a few properties.
First, when given a positive budget, \search should either finish the given job
or return a list of unexplored nodes.  Any unexplored node should
represent a smaller portion of the unfinished work, i.e.\ running \search 
(with positive budgets) on the unexplored nodes and any resulting unexplored nodes will
eventually result in
finishing the original job.  The code should also interpret the budget in some suitable
way where larger budgets correspond to doing more work than smaller budgets.  
This may require some modification of the legacy code.
Our applications usually interpret the budget as \emph{number of traversed nodes}
and \emph{depth}, but this is not required (see conflict budgeting in Section~\ref{subsec:mtsat}).

Any given worker must be able to work on any given unexplored node that \mts has seen.
It is helpful for the unexplored nodes to represent non-overlapping jobs.
\mts supports sharing data between workers, but it is helpful for shared data
to be small. Implementing a
shared memory version of \mts could help performance when large
amounts of data are shared.
Shared data is not used in our enumeration applications.  It is used for satisfiability
and similar applications to prune the search tree.

\subsection{Master process}
\label{subsec:mts_master}

The master process begins with initialization, including obtaining
an application-provided initial \startvertex.  
It places this initial subproblem in a (new) job list $L$, and then
enters the main loop.
In this main loop, the master assigns budgeted subproblems to workers, collects unfinished subproblems to add to $L$, 
and collects/sends updated \shared from/to the workers.
Assigning \shared updates to the master is not essential: it simplifies
checkpointing but can increase load on the master and interconnect.
Each worker either finishes its subproblem or reaches its budget limitation (\mymaxdepth and \maxnodes)
and returns unfinished subproblems to the master for insertion into $L$.
This continues until
no workers are running and the master has no unfinished subproblems.
Once the main loop ends, the master informs all processes to finish.
The main loop performs the following tasks:
\begin{itemize}
 \item subproblems and relevant \shared updates are sent to free workers when
       available;
 \item check if any workers are done, mark them as free and receive 
       their unfinished subproblems;
 \item check and receive \shared updates.
\end{itemize}
Pseudocode is given as Algorithm \ref{alg:mts_master} in the Appendix.
Communication is non-blocking and work proceeds when required information
is available.

Using reasonable parameters is critical to performance.
This is done dynamically
by observing $|L|$. We use parameters \lmin, \lmax and \myscale which
depend on the type of tree search problem being handled.
The following default values are used in this paper.
Initially, to create a reasonable size list $L$, we set  $\mymaxdepth=2$ and $\maxnodes=5000$.
Therefore the initial worker will generate subtrees at depth 2 until 5000 nodes have
been visited and then terminates sending roots of unvisited subtrees back to the master. 
Additional workers are given the same aggressive parameters
until $|L|$ grows larger than \lmin times the number of processors, at which point \mymaxdepth
is removed. Once $|L|$ is larger than \lmax times the number of processors,
we multiply
the budget by \myscale. With $\myscale=40$ 
workers will
not generate any new subproblems unless their tree has at least 200,000 nodes.
If $|L|$ drops below these bounds we return to the smaller budgets.
The default is $\lmin=1,\lmax=3$.
In Section~\ref{sec:hist} we show an example of how $|L|$
typically behaves with these settings.

\subsection{Workers}

The worker processes are simpler -- they receive the problem at
startup, and then repeat their main loop: receive a
parametrized subproblem and possible \shared updates from the master,
work on the subproblem subject to the
parameters, send the output to the consumer, and send updated
\shared and unfinished subproblems to the master if the budget is exhausted.  
Pseudocode is given as Algorithm \ref{alg:mts_worker} in the Appendix.

\subsection{Consumer process}

The consumer process in \mts is the simplest.  The workers send output
to the consumer in exactly the format it should be output (i.e., this
formatting is done in parallel).  The consumer simply outputs it.
By synchronizing output to a single destination, the consumer 
delivers a continuous output stream to the user in the same way as 
\search does. Pseudocode is given 
as Algorithm \ref{alg:mts_consumer} in the Appendix.

\section{Applying \mts to reverse search}
\label{sec:app1}

Reverse search is a technique for generating large relatively unstructured sets of discrete
objects~\cite{AF93}. 
In its most basic form, reverse search can be viewed as the traversal of a spanning tree, called the reverse
search tree $T$, of a graph $G=(V,E)$ whose nodes are the objects to be generated. Edges in the graph are
specified by an adjacency oracle, and the subset of edges of the reverse search tree are
determined by an auxiliary function, which can be thought of as a local search function $f$ for an
optimization problem defined on the set of objects to be generated. One vertex, $v^*$, is designated
as the {\em target} vertex. For every other vertex $v \in V$ 
repeated application of $f$ must generate a
path in $G$ from $v$ to $v^*$. The set of these paths defines the reverse search tree $T$, which has root $v^*$.

A reverse search is initiated at $v^*$, and only edges of the reverse search tree are traversed.
When a node is visited, the corresponding object is output.  Since there is no possibility of
visiting a node by different paths, the visited nodes do not
need to be stored.  Backtracking can be performed in the
standard way using a stack, but this is not required as the local search function can be used for
this purpose. 

In the basic setting described here a few properties are required. Firstly, the
underlying graph $G$ must be connected and an upper bound on the maximum vertex degree, $\Delta$, must
be known.  The performance of the method depends on $G$ having $\Delta$ as low as
possible.  An adjacency oracle $\Adj(v,j)$ must be capable of generating
the adjacent vertices of any given
vertex $v$ in $G$.
For each vertex $v \neq v^*$
the local search function $f(v)$ returns the tuple $(u,j)$ where $v = \Adj(u,j)$ which defines the parent $u$
of $v$ in $T$.
Pseudocode is given in Algorithm~\ref{rsalg1} and is invoked by
setting $\startvertex = v^*$. C implementations for
several simple enumeration problems are given at~\cite{tutorial}.
For convenience later, we do not output the $\startvertex$ in the pseudocode shown.
Note that the vertices are output as a continuous stream.
Also note that Algorithm~\ref{rsalg1} does not
require the parameter $\startvertex$ to be the root $v^*$ of the entire search tree. If
an arbitrary node in the tree is given, the algorithm reports the subtree
rooted at this node and terminates.

We need to implement budgeting in order to parallelize
Algorithm~\ref{rsalg1} with \mts.
We do this in two ways that may be combined.
Firstly we introduce the parameter \mymaxdepth which terminates the
tree search at that depth returning any unvisited subtrees.
Secondly we introduce a parameter \maxnodes which terminates the tree search
after this many nodes have been visited and again returns the
roots of all unvisited subtrees.
This entails backtracking to the root
and returning the unvisited siblings of each node in the backtrack path.
These modifications are straightforward and given in Algorithm~\ref{bts},
which reduces to Algorithm~\ref{rsalg1}
by deleting the items in red.

\begin{figure*}[htb]
\noindent
\begin{minipage}[t]{0.42\textwidth}
\begin{algorithm}[H]
\begin{algorithmic}[1]
\Procedure{rs}{$\startvertex$}
        \State $v \gets \startvertex$,~$j \gets 0$,~$\mydepth \gets 0$
        \Repeat
 \Statex
       	\While {$j < \Delta$}
                \State $j \gets j+1$
		\If {$f(\Adj(v,j)) = v$}  
			\State $v \gets \Adj(v,j)~~~~~$  
			\State $j \gets 0$ 
\Statex
                        \State $\mydepth \gets \mydepth+1$         
\Statex
\Statex
\Statex
\Statex
			\State \putoutput ($v$)         
                \EndIf
        \EndWhile
        \If {$\mydepth > 0$}   
		\State $(v,j) \gets f(v)$
                \State $\mydepth \gets \mydepth-1  $       
        \EndIf
        \Until {$\mydepth=0$ {\bf and} $j=\Delta$}
\EndProcedure
\end{algorithmic}
\caption{Generic Reverse Search}
\label{rsalg1}
\end{algorithm}
\end{minipage}
\begin{minipage}[t]{0.58\textwidth}
\begin{algorithm}[H]
\begin{algorithmic}[1]
\Procedure{brs}{$\startvertex$, \textcolor{red}{$\mymaxdepth$, $\maxnodes$}}
        \State $j \gets 0~~~v \gets \startvertex~~~\mydepth \gets 0$ \textcolor{red}{$~~~ \mycount \gets 0 $}
        \Repeat
        \State \textcolor{red}{$\unexplored \gets \myfalse$}
        \While {$j < \Delta$ \textcolor{red}{{\bf and} $\unexplored = \myfalse$} }
                \State $j \gets j+1$
                \If {$f(\Adj(v,j)) = v$}  \Comment{forward step}
                        \State $v \gets \Adj(v,j)$
                        \State $j \gets 0$
                        \textcolor{red}{\State $\mycount \gets \mycount+1$}
                        \State $\mydepth \gets \mydepth + 1$
                        \color{red}
                        \If  {$\mycount \ge \maxnodes$ {\bf or}
                          \Statex\hspace{6em}\hphantom{\textbf{if}~}$\mydepth = \mymaxdepth$}
                            \State $\unexplored \gets \mytrue$ \Comment{over budget}
                        \EndIf
                        \color{black}
                        \State \putoutput $(v,\textcolor{red}{\unexplored})$
                \EndIf
        \EndWhile
        \If {$\mydepth > 0$}   \Comment{backtrack step}
                \State $(v,j) \gets f(v)$
                \State $\mydepth \gets \mydepth - 1$
        \EndIf
        \Until {$\mydepth = 0$ {\bf and} $j=\Delta$}
\EndProcedure
\end{algorithmic}
\caption{Budgeted Reverse Search}
\label{bts}
\end{algorithm}
\end{minipage}
\end{figure*}

To output all nodes in the subtree of $T$ rooted at $v$ we set  
$\startvertex=v$, $\maxnodes=+\infty$ and $\mymaxdepth=+\infty$.
To break up $T$ into subtrees we have two options that can be combined.
Firstly we can set the \mymaxdepth parameter resulting in all nodes at
that depth to be flagged as unexplored.
Secondly we can set the budget parameter \maxnodes.
In this case, once this many nodes have been explored the current node
and all unexplored siblings on the backtrack path to the root are output
and flagged as unexplored.

\subsection{Example 1: Topological sorts}
\label{sec:mtop}
A C implementation ({\em per.c})
of the reverse search algorithm for generating permutations is
given in the tutorial~\cite{tutorial}.
A small modification of this code generates all
topological sorts of a partially ordered set that is given by
a directed acyclic graph (DAG). Such topological sorts are also called linear extensions 
or topological orderings.
The code modification is given as Exercise 5.1 and a solution to
the exercise 
({\em topsorts.c}) is at~\cite{tutorial}.
Here we describe how to modify this code to allow
parallelization via the \mts interface to produce the program \mtop.
The details and code are available at \cite{tutorial}.

It is convenient to describe the procedure as two phases. Phase
1 implements budgeting and organizes the internal data in a suitable
way. This involves modifying an implementation
of Algorithm~\ref{rsalg1} to an implementation of Algorithm~\ref{bts}
that can be independently tested. We need to prepare a global data structure  bts\_data
which contains problem data obtained from the input.
In Phase 2 we build a node structure for use by the \mts wrapper and add necessary
routines to allow initialization and I/O in a parallel setting. In practice
this involves using a header file from \mts. The resulting program
{\em btopsorts.c} can be compiled as a sequential code or with \mts
as a parallel code with no change in the 
source files.

In the second phase we add the `hooks' that allow communication with \mts.
This involves defining a Node structure which holds all necessary information
about a node in the search tree. The roots of unexplored subtrees are maintained
by \mts for parallel processing. Therefore whenever a search terminates due
to the \maxnodes or \mymaxdepth restrictions, the Node structure of each unexplored
tree node is returned to \mts. As we do not wish to customize \mts for
each application, we use a very generic node structure. The user should pack
and unpack the necessary data into this structure as required. The Node
structure is defined in the \mts header.

The efficiency of \mts depends on keeping the job list non-empty
until the end of the computation, without letting it get too large. Depending on
the application, there may be a substantial restart cost for each unexplored
subtree. Surely there is no need to return a leaf as an unexplored node, and the
{\em prune=0} option checks for this. Further, if an unexplored node has only
one child it may be advantageous to explore further, terminating either at
a leaf or at a node with two or more children, which is returned
as {\em unexplored}. The {\em prune=1}
option handles this condition, meaning that no isolated nodes or paths are
returned as unexplored.
Note that pruning is not a built-in \mts option; it is an example of options
that applications may wish to include and was implemented in \mtop.

\subsection{Example 2: Spanning trees}
\label{sec:tree}
In the tutorial~\cite{tutorial} a C implementation ({\em tree.c}) is given
for the reverse search algorithm for all spanning trees of the complete graph.
An extension of this to generate all spanning trees of 
a given graph is stated as Exercise 6.3.
Applying Phase 1 and 2 as described above results in the code
{\em btree.c}. Again this may be compiled as a sequential code or with
the \mts wrapper 
to provide the parallel implementation \mtree.
All of these codes are given at the URL~\cite{tutorial}.

\subsection{Example 3: Triangulations}
\label{sec:triang}

The previous examples are from the tutorial on reverse search; a more
substantial application of \mts is the parallel enumeration of 
triangulations implemented in \mptopcom~\cite{JJK17}.  This was the first
parallel code for the problem, and has already found
applications~\cite{JK18,Sch17}.
Here we give a brief summary, see~\cite{JJK17} for details of \mptopcom
and~\cite{triang} for background on triangulations.

Enumerating triangulations in the plane is one of the early
applications of reverse search~\cite{AF93}.  Imai et al.~\cite{Imai02}
later gave a reverse search algorithm for enumerating triangulations in
general dimensions, including techniques for restricting to regular
triangulations and enumerating up to symmetry.  However, the standard
tool used for this is \topcom~\cite{TOPCOM},
which essentially performs a breadth-first search of the flip graph.

\mptopcom is an implementation of the algorithm by Imai et al.~\cite{Imai02}
with some improvements, parallelized using \mts.  It uses \topcom for 
triangulations and flips and \polymake~\cite{polymake} for basic data types.
The single-threaded version is competitive with other codes for this problem,
while the parallel version allows one to enumerate triangulations that are 
(in practice) beyond reach of the other codes; see~\cite{JJK17} for
detailed experimental results.

\subsection{Example 4: Galton Watson trees}
\label{sec:GW}

The previous examples give reverse search trees that are rather well balanced.
By that we mean that the height of the tree is either logarithmic or 
polylogarithmic in its size, i.e. the number of nodes in the tree. Balanced trees are
especially suited to parallelization and this is indeed the case
for \mts as we will see in the experimental results 
in Section \ref{sec:comp}. To see how well budgeting works for
non-balanced trees Avis and Devroye \cite{AD17a} studied
its behaviour on a family of random trees with depth roughly the square root
of their size. They also studied how the job list size $L$ depends on the
budget parameter $b$. We review the main result.

A Galton-Watson (or Galton-Watson-Bienaym\'e) tree
is a rooted random ordered tree that is a classical statistical model of a birth and death process. 
Each node independently
generates a random number of children drawn from a fixed offspring distribution $\xi$.
The distribution of $\xi$ defines the distribution of $T$,
a random Galton-Watson tree.
They consider critical Galton-Watson trees which are those
having $\EXP \{ \xi = 1 \}$, and $\PROB \{ \xi = 1 \} < 1$.
In addition, they assume that the variance of $\xi$ is finite (and hence,
nonzero).

Moon \cite{moon70} and Meir and Moon \cite{MM78} defined
the {\it simply generated trees} as ordered labelled trees
of size $n$ that are all equally likely given
a certain pattern of labeling for each node of a given degree.
The most important examples include the Catalan trees (equiprobable
binary trees), the equiprobable $k$-ary trees,
equiprobable unary-binary trees (ordered trees with up to two children),
random Motzkin  trees,
random planted plane trees (equiprobable
ordered trees of unlimited degrees) and Cayley trees
(equiprobable unordered rooted trees).
It turns out that all these trees can be represented
as critical Galton-Watson trees conditional on their size, $n$,
a fact first pointed out by Kennedy \cite{kennedy75}
and further developed by others.
For example, when
$\xi$ is $0$ or $2$ with probability $1/4$, and $1$ with probability $1/2$,
we obtain the uniform binary (Catalan) tree.
Uniformly random full binary trees are obtained by setting
$\PROB \{ \xi = 0 \} = \PROB \{ \xi = 2 \} = 1/2$.
A uniformly random $k$-ary tree has its offspring distributed
as a binomial $(k, 1/k)$ random variable.
A uniform planted plane tree is obtained for the geometric law
$\PROB \{ \xi = i \} = 1/2^{i+1}$,
$i \ge 0$.
When $\xi$ is Poisson of parameter $1$, one obtains
(the shape of) a random rooted labeled (or Cayley) tree.
For $\xi$ uniform on $\{ 0, 1 , 2, \ldots, k  \}$, $T_n$ is like
a uniform ordered tree with maximal degree of $k$.
All such trees can be dealt with at once in the
Galton-Watson framework.

In \cite{AD17a} an analysis is performed on the expected size
of the job list as a function of the budget parameter $b$.
They prove that if $T_n$ is a Galton-Watson tree of size $n$ determined by
$\xi$, where $\EXP \{ \xi \} = 1$ variance $0 <  \sigma^2 < \infty$,
then
\begin{equation}
\label{bbound}
{L_n \over n} \to \sqrt{\frac{\pi \sigma^2}{8b} }
\end{equation}
in probability as $n \to \infty$, where
$b$ is the budget and $L_n$ is the job list size.
For example, for the Catalan trees we have $\sigma^2=3/2$
and so the constant on the right hand side of (\ref{bbound})
is $\sqrt{3 \pi/16b}$. With a budget $b=5000$ this indicates
that about 1\% of the nodes are returned unexplored to the job list.
Jobs returned to $L_n$ are the main source of overhead in \mts, so a value of about
1\% is very satisfactory in practice.

Experimental results are given in \cite{AD17a} for various Galton-Watson trees to show that the estimate 
in (\ref{bbound}) is
quite accurate. 

\section{Applying \mts to satisfiability}
\label{sec:sat}
Boolean satisfiability (SAT) asks us to determine the existence of
(or find) satisfying assignments for propositional formulas, 
see~\cite{HandbookOfSAT2009} for more background.  SAT solvers
have made tremendous progress over the years, and are now widely used as general
NP solvers.  While most application problems seem to result in easy SAT
instances~\cite{hot17}, there has long been interest in parallel SAT solvers
for hard instances.  Despite the many challenges~\cite{HW12, satbarriers}
in parallel SAT, there are recent successes~\cite{HKM16}.

There are two major approaches to parallel SAT solvers.  Either one
somehow partitions the space of possible assignments and uses
divide-and-conquer (e.g.,~\cite{ALST16} for a recent example) or
one uses the portfolio approach and runs many sequential
solvers on the original problem (e.g., \plingeling~\cite{Biere12}).
In either case, a major issue is determining which learnt 
clauses\footnote{CDCL solvers learn clauses during the
search, pruning the search space.
See, e.g., Chapter 4 of~\cite{HandbookOfSAT2009}.} to share between 
workers~\cite{AS14}.  While sharing these clauses helps prune the
search space, additional clauses slow the solver and enormous numbers of
clauses are learned.

Another question for divide-and-conquer solvers is the question of
how to divide the search space.  Many approaches have been tried,
often setting initial variables and using a common feature of 
sequential solvers to ``solve under assumptions''.  
Some recent solvers (e.g.,~\cite{ALST16} and \treengeling~\cite{Biere12})
work on these subproblems subject to some budget, and hard subproblems 
can be split again.
Cube-and-conquer~\cite{cubeconquer} is another recent approach that uses
look-ahead solvers to divide the search space for CDCL solvers.

\subsection{\mtsat: parallelizing \minisat with \mts}
\label{subsec:mtsat}

We used \mts to implement a divide-and-conquer solver
\mtsat, using \minisat 2.2.0 as sequential solver.
Our goal was to demonstrate the use of \shared and
show that \mts can be used in settings other than enumeration.
\mtsat is still experimental and much work remains to reach
the level of state-of-the-art dedicated parallel SAT solvers,
but it allows for experimentation with, e.g., budgeting and
restart strategies in parallel SAT.

\minisat~\cite{minisat} supports solving under assumptions,
i.e.\ solving subject to some partial assignment.  It also 
supports solving subject to a budget,
given in propagations or conflicts, returning
\emph{unknown} if the given subproblem
could not be solved within the budget.

The major modification required is to report unexplored
partial assignments when the budget is exhausted.  At any point in
the search, SAT solvers distinguish between decision variables
and propagated variables.  Decision variables are those where the
solver chose an assignment, while propagated variables are those
where the solver was able to determine (because of a unit clause) that
only one option need be explored.  It suffices to return
unexplored nodes corresponding to the current partial assignment and
to those formed by taking the unexplored options for decision variables
(including the last one) along the backtrack path.

Regarding learnt clauses,
we implemented a simple scheme sharing only learnt unit clauses. The 
idea is that short clauses cut the search tree more
than longer clauses; an early version of \plingeling also
shared only units~\cite{Biere12}.  We avoided more sophisticated
approaches to sharing clauses~\cite{AS14,ALST16}, using conflicts
to prune the job list $L$ and similar ideas for simplicity.

\mtsat includes additional options.
For example, while the parallel solvers most similar to
our approach~\cite{ALST16,Biere12} budget using conflicts -- we
added the option to budget using \emph{decisions}.
Conflict budgets correspond to hitting a leaf in the search
tree, while decision budgets correspond to nodes in the search space
(omitting propagated variables since those are forced).
Conflict budgets are attractive, but decision budgets correspond 
more closely to the budgets used in Section~\ref{sec:app1} and
allow us to experiment with different budgeting techniques.

Modern solvers generally perform random
restarts, abandoning the current search to start
over (cf. Chapter 4 of~\cite{HandbookOfSAT2009}) and hopefully avoid
getting stuck in hard parts of the search space.  We
split problems along the backtrack path and schedule these abandoned
portions of the search space for later exploration -- possibly resulting
in much duplicated work.
We therefore added an option to disable restarts, in
order to experiment with their impact on performance in \mtsat,
and formula preprocessing, to experiment with the idea that
avoiding preprocessing can be beneficial to divide-and-conquer parallel
SAT solvers~\cite{HW12}.

The total is 50 lines of changes to legacy \minisat (including
support to parse inputs from strings) of the original 4803 lines,
plus a few hundred lines of generic code interfacing the \minisat API
and \mts that can be re-used.
Essentially identical changes suffice to parallelize \glucose
(since it is based on \minisat) and others, and so we also parallelize
\glucose 3.0.
One could easily support 
workers using a mix of solvers, a hybrid of the divide-and-conquer and
portfolio approaches to parallel SAT.

\section{Experimental results}
\label{sec:comp}

The tests were performed at Kyoto University on \maic, a cluster of 5 nodes 
with a total of 192 identical processor cores, consisting of: 
\begin{itemize}
\item
\mainew: 4 nodes, each containing: 2x Opteron 6376 (16-core 2.3GHz), 32GB memory, 500GB hard drive (128 cores in total);
\item
\mainewer: 4x Opteron 6376 (16-core 2.3GHz), 64 cores, 256GB memory, 4TB hard drive.
\end{itemize}
A complete description of the problems solved below is given in \cite{AJ16a} and the input files
are available by following the link to tutorial2 at \cite{tutorial}.

\subsection{Topological sorts: \mtop}
\label{subsec:mtop}

The tests were performed using the following codes:
\begin{itemize}
\item
\vr: obtained from~\cite{COS}, generates topological sorts
in lexicographic order via the     
Varol-Rotem algorithm~\cite{VR81} (Algorithm V in Section 7.2.1.2 of~\cite{knuth11});
\item
\genle: also obtained from~\cite{COS}, generates topological sorts in Gray code order
using the algorithm of Pruesse and Rotem~\cite{PR91};
\item
\btop: derived from the reverse search code {\em topsorts.c}~\cite{tutorial} as 
described in Section~\ref{sec:mtop};
\item
\mtop: \mts parallelization of \btop.
\end{itemize}
For the tests all codes were used in count-only mode due to the enormous output that 
would otherwise be generated. All codes were used with default parameters:
\begin{equation}
\mymaxdepth=2~~ \maxnodes=5000~~ \myscale=40~~ \lmin=1~~ \lmax=3
\label{defpar}
\end{equation}

The following graphs were chosen, listed in order of increasing
edge density: \pmtwotwo, \catfourtwo, \keightnine.
The constructions for the first two partial orders
are well known (see, e.g., Section 7.2.1.2 of~\cite{knuth11}) and the third is 
a complete bipartite graph.

\begin{table}[htbp]
\centering
\scalebox{0.88}{
\begin{tabular}{||c c c c||c|c|c||c|c|c|c|c||}
 \hline

Graph &m&n   & No. of perms  &\vr& \genle  & \btop  &\multicolumn{5}{|c||}{\mtop  }   \\
      &nodes & edges    &   & &   &  & 12 & 24 & 48 & 96 & 192  \\
\hline
\pmtwotwo &22&21    & 13,749,310,575  & 179  & 14 & 12723  & 1172 &595& 360&206 &125 \\
\hline
\catfourtwo &42 &61   & 24,466,267,020  & 654  & 171 &45674  &4731 &2699&1293&724 &408 \\
\hline
\keightnine &17&72    & 14,631,321,600  & 159  & 5 & 8957 &859&445&249 & 137   &85  \\
\hline
\end{tabular}
}
\caption{Topological sorts: \maic, times in secs} 
\label{tab:tops}
\end{table}

Results are in Table~\ref{tab:tops}. The reverse search
code \btop is very slow, over 900 times slower than \genle and
over 70 times slower than \vr on \pmtwotwo. However     
the parallel \mts code obtains excellent speedups and is faster than \vr on all problems when
192 cores are used. 

\subsection{Spanning trees: \mtree}

The tests were performed using the following codes:
\begin{itemize}
\item
\gray: Knuth's implementation~\cite{knuthcode} of an algorithm that generates all spanning trees of a given graph,
changing only one edge at a time, as described in
Malcolm Smith's M.S. thesis, {\em Generating spanning trees} (University
of Victoria, 1997);
\item
\graysp: Knuth's improved implementation of \gray:
``This program combines the ideas of \gray
and \progname{spspan}, resulting in a glorious routine that generates
all spanning trees of a given graph, changing only one edge at a time,
with `guaranteed efficiency'---in the sense that the total running
time is $O(m+n+t)$ when there are $m$ edges, $n$ vertices, and $t$
spanning trees.''~\cite{knuthcode};
\item
\btree: derived from the reverse search code {\em tree.c}~\cite{tutorial} as described in Section~\ref{sec:tree};
\item
\mtree: \mts parallelization of \btree.
\end{itemize}
\noindent
Both \gray and \graysp are described in detail in Knuth~\cite{knuth11}.
Again all codes were used in count-only mode 
and with the default parameters (\ref{defpar}).
The problems chosen were the following graphs which are listed in order of increasing
edge density: \eightcage, \pfivecfive, \cfivecfive, \ksevenseven, \konetwo. 
The latter 4 graphs were motivated by Table 5 in~\cite{knuth11}: \pfivecfive appears therein
and the other graphs are larger versions of examples in that table.

\begin{table}[h!tbp]
\centering
\scalebox{0.9}{
\begin{tabular}{||c c c c||c|c|c||c|c|c|c|c||}
 \hline

Graph &m&n   & No. of trees  &\gray& \graysp & \btree &\multicolumn{5}{|c||}{\mtree  }   \\
 &nodes & edges &   & &  &  & 12 & 24 & 48 & 96 & 192  \\
\hline
\eightcage &30&45    &  23,066,015,625   & 3166 & 730 & 10008 &1061 &459& 238&137 & 92  \\
\hline
\pfivecfive  & 25& 45 & 38,720,000,000  & 3962 & 1212& 8918  & 851 &455& 221&137 & 122 \\
\hline
\cfivecfive  & 25 & 50 &1,562,500,000,000  & 131092 & 41568&230077    &26790 &13280&7459& 4960 & 4244 \\
\hline
\ksevenseven & 14 & 49 &  13,841,287,201   &  699 & 460 &  2708 &259  &142&  68& 51 & 61 \\
\hline
\konetwo & 12 & 66 &  61,917,364,224   & 2394 & 1978 & 3179 &310 &172&  84& 97 & 148 \\
\hline
\end{tabular}
}
\caption{Spanning tree generation: \maic, times in secs} 
\label{tab:trees}
\end{table}
The computational results are given in Table~\ref{tab:trees}. This time the reverse search
code is a bit more competitive:  about 3 times slower than \gray and
about 14 times slower than \graysp on {\em 8-cage} for example. 
The parallel \mts code runs about as fast as \graysp on all problems when
12 cores are used and is significantly faster after that. Near linear speedups are
obtained up to 48-cores but then tail off. For the two dense graphs
$ K_{7,7}$ and $K_{12}$ the performance of \mts is actually worse with 192 cores than with 96. 

\subsection{Satisfiability}
\label{subsec:satexp}

The tests were performed using the following codes:
\begin{itemize}
 \item \minisat: version 2.2.0, classic sequential solver~\cite{minisat};
 \item \glucose: version 3.0, sequential solver~\cite{glucose} derived from \minisat;
 \item \mtsat: parallel solver using \mts and \minisat 2.2.0;
 \item \mtsatglucose: parallel solver using \mts and \glucose 3.0;
 \item \glucosesyrup: version 4.0, parallel (shared memory) solver;
 \item \lingeling, \treengeling: version bbc, sequential and (shared memory) parallel solvers~\cite{Biere12}.
\end{itemize}

Benchmarking parallel SAT solvers is
challenging~\cite{HW12} and any particular instance may give
 superlinear speedups or timeouts.  We use a standard set of hard
instances from applications, and count the number of problems that each solver
can solve within a given time.  We re-use the
setup of~\cite{ALST16}, i.e. the 100 instances in the parallel
track of SAT Race 2015~\cite{satrace2015} and a timeout of 20 minutes.
Results are in Figure~\ref{fig:mtsatperf}.
Due to different computers used, our results are not
directly comparable to those in~\cite{ALST16}.  As noted by~\cite{hot17},
solvers like \mtsat can use substantial memory on very large instances,
limiting the number of processes that can execute in a given amount of memory.
The computers we used had sufficient memory for the instances used.

\begin{figure}[h!tb]
\centering
\begin{subfigure}{0.49\textwidth}
 \centering
 \resizebox{\textwidth}{!}{ \input{plots/cactus-sat.tex} }
\caption{Instances solved vs time}
\end{subfigure}
\begin{subfigure}{0.49\textwidth}
 \centering
 \begin{tabular}{|r|c|c|c|}
  \hline
  Solver      &SAT&UNSAT&Total\\
  \hline
  \minisat    &18 &  1  &  19 \\
  \mtsat (16) &17 &  1  &  18 \\
  \mtsat (32) &22 &  2  &  24 \\
  \mtsat (64) &23 &  4  &  27 \\
  \mtsat (128)&29 &  7  &  36 \\
  \mtsat (192)&35 & 10  &  45 \\
  \hline
  \lingeling  & 17 & 10 & 27 \\
  \treengeling (32) & 38 & 21 & 59 \\
\hline
\end{tabular}
\caption{Instances solved within 1200s}
\end{subfigure}
\caption{\mtsat performance (decision budgeting, default parameters (\ref{defpar}))}
\label{fig:mtsatperf}
\end{figure}

The results in Figure~\ref{fig:mtsatperf} show
improvement from additional cores using default parameters
and decision budgeting with no attempt at tuning.  Performance with conflict
budgeting
is shown in Figure~\ref{fig:mtsatperfconfbudg}, using an initial budget of
$10000$ conflicts (i.e. the corresponding value in~\cite{ALST16}).

\begin{figure}[htb]
\centering
\begin{subfigure}{0.49\textwidth}
 \centering
 \resizebox{\textwidth}{!}{ \input{plots/cactus-sat-confbudg.tex} }
\caption{Instances solved vs time}
\end{subfigure}
\begin{subfigure}{0.49\textwidth}
 \centering
 \begin{tabular}{|r|c|c|c|}
  \hline
  Solver      &SAT&UNSAT&Total\\
  \hline
  \minisat    &18 &  1  &  19 \\
  \mtsat (16) &18 &  2  &  20 \\
  \mtsat (32) &23 &  3  &  26 \\
  \mtsat (64) &27 &  7  &  34 \\
  \mtsat (128)&30 &  10 &  40 \\
  \mtsat (192)&34 &  11 &  45 \\
\hline
\end{tabular}
\caption{Instances solved within 1200s}
\end{subfigure}
\caption{\mtsat performance (conflict budgeting, $\maxnodes=10000$, $\myscale=10$)}
\label{fig:mtsatperfconfbudg}
\end{figure}

All non-timeout outputs are correct, and the 32-core run with conflict budgeting
solves problem \texttt{62bits\_10.dimacs.cnf} (reported as unsolved in the SAT Race 2015
results) giving a correct satisfying assignment.  It is likely that experimenting
with parameter values can improve performance, and using a newer sequential
solver on the workers may be another source of improvement given the
performance \treengeling achieves starting from the higher baseline performance
of \lingeling.

Along these lines, we also report results applying \mts to parallelize \glucose.  
Figure \ref{fig:mtsatglucoseperf} shows results using the default
decision budgeting and Figure \ref{fig:mtsatglucoseperfconfbudg} shows
results using conflict budgeting.  Note that like in \mtsat, only
unit clauses are shared in \mtsatglucose.  A more sophisticated approach
to sharing learnt clauses, e.g. sharing glue clauses,
would likely help performance. 

\begin{figure}[htb]
\centering
\begin{subfigure}{0.49\textwidth}
 \centering
 \resizebox{\textwidth}{!}{ \input{plots/cactus-sat-glucose.tex} }
\caption{Instances solved vs time}
\end{subfigure}
\begin{subfigure}{0.49\textwidth}
 \centering
 \begin{tabular}{|r|c|c|c|}
  \hline
  Solver      &SAT&UNSAT&Total\\
  \hline
  \glucose           & 6 &  5  & 11 \\
  \mtsatglucose (16) &26 &  5  & 31 \\
  \mtsatglucose (32) &26 &  5  & 31 \\
  \mtsatglucose (64) &31 &  6  & 37 \\
  \mtsatglucose (128)&32 &  9  & 41 \\
  \mtsatglucose (192)&35 & 10  & 45 \\
\hline
  \glucosesyrup (32) &32 & 18  & 50 \\
  \glucosesyrup (64) &31 & 18  & 49 \\
\hline
\end{tabular}
\caption{Instances solved within 1200s}
\end{subfigure}
\caption{\mtsatglucose performance (decision budgeting, default parameters (\ref{defpar}))}
\label{fig:mtsatglucoseperf}
\end{figure}

\begin{figure}[htb]
\centering
\begin{subfigure}{0.49\textwidth}
 \centering
 \resizebox{\textwidth}{!}{ \input{plots/cactus-sat-glucose-confbudg.tex} }
\caption{Instances solved vs time}
\end{subfigure}
\begin{subfigure}{0.49\textwidth}
 \centering
 \begin{tabular}{|r|c|c|c|}
  \hline
  Solver      &SAT&UNSAT&Total\\
  \hline
  \glucose           & 6 &  5  &  11 \\
  \mtsatglucose (16) &23 &  7  &  30 \\
  \mtsatglucose (32) &27 &  7  &  34 \\
  \mtsatglucose (64) &30 & 10  &  40 \\
  \mtsatglucose (128)&34 & 13  &  47 \\
  \mtsatglucose (192)&38 & 13  &  51 \\
\hline
\end{tabular}
\caption{Instances solved within 1200s}
\end{subfigure}
\caption{\mtsatglucose performance (conflict budgeting, $\maxnodes=10000$, $\myscale=10$)}
\label{fig:mtsatglucoseperfconfbudg}
\end{figure}

In all cases, we see that additional cores allow one to solve more problems given a fixed amount of time.
While it is likely that performance can be improved by tuning and a better approach to sharing clauses,
these results suffice for our purpose: to show that one can easily parallelize a legacy code with \mts.

As mentioned earlier, conflict budgeting is most common in this kind of parallel solver.  This
is because generating a conflict clause guarantees at least some progress has been made and instances
can have very different and enormous numbers of variables.  Conflict budgeting usually slightly
outperformed decision budgeting in the runs here, however this was not universal.  Given the minimal
tuning for both budget types, our results do not show a clear difference regarding how to budget.

\section{Evaluating and improving performance}
\label{sec:hist}

\begin{figure}[htb]
\centering
\begin{subfigure}[b]{0.49\textwidth}
 \centering
 \resizebox{\textwidth}{!}{ \input{plots/efficiency-mtopsorts.tex} }
 \caption{Efficiency with \mtop}
 \label{subfig:mtopefficiency}
\end{subfigure}
\begin{subfigure}[b]{0.49\textwidth}
 \centering
 \resizebox{\textwidth}{!}{ \input{plots/efficiency-mtree.tex} }
 \caption{Efficiency with \mtree}
 \label{subfig:mtreeefficiency}
\end{subfigure}
\caption{Efficiency vs number of cores (data from Tables \ref{tab:tops} and \ref{tab:trees})}
\label{fig:efficiency}
\end{figure}
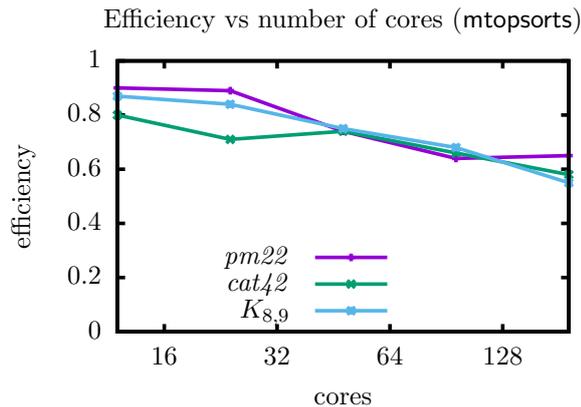
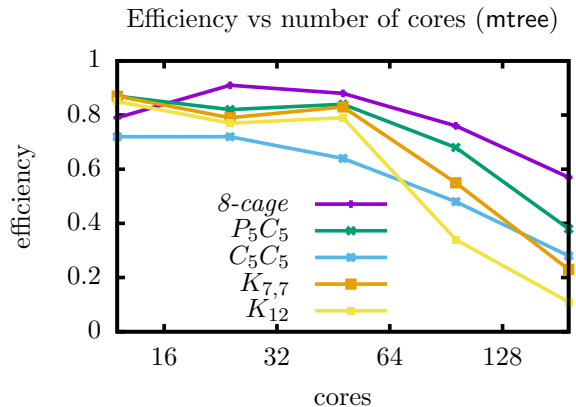

Our main measures of performance for the enumeration problems are the elapsed time taken and
the {\em efficiency} defined as:
\begin{equation}
\mathrm{efficiency ~=~ \frac{single~core~running~time}{number~of~cores * multicore~running~time}}
\label{eff}
\end{equation}
Multiplying efficiency by the number of cores gives the speedup.
Speedups that scale linearly with the number of cores give constant efficiency.
External factors can affect performance as the load on the machine increases.  One
example is dynamic overclocking, where the speed of working cores may be increased by
25\%--30\% when other cores are idle.  This limits the maximum efficiency achievable when all cores
are used, since the single core running times are measured on otherwise idle machines.
In Figure \ref{fig:efficiency} we plot the efficiencies obtained by 
\mtop and \mtree for the runs shown in Tables \ref{tab:tops} and \ref{tab:trees} respectively. 

\begin{figure}[h!tbp]
\begin{minipage}{0.5\textwidth}
 \includegraphics[width=\textwidth]{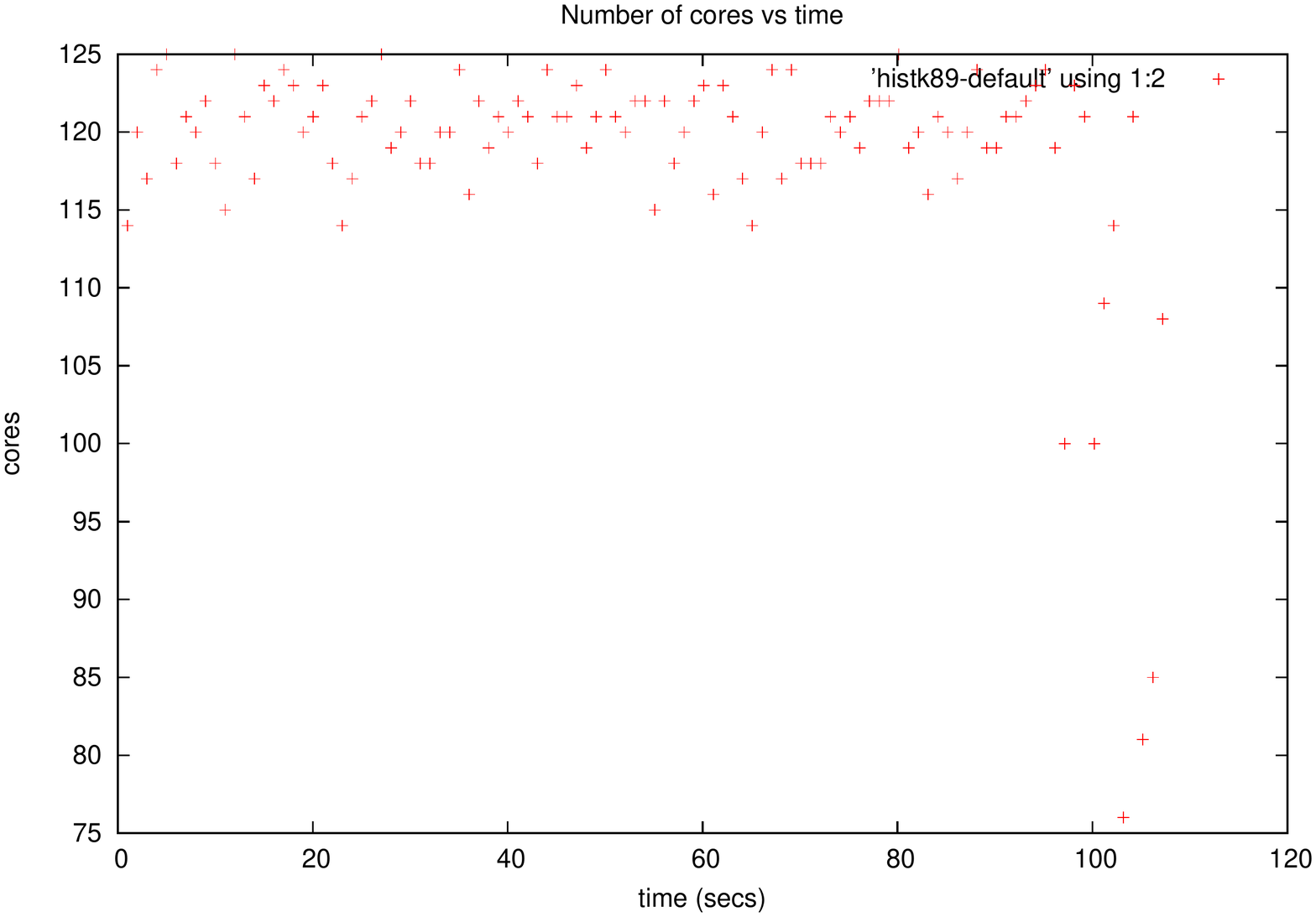}
\end{minipage}
\begin{minipage}{0.5\textwidth}
 \includegraphics[width=\textwidth]{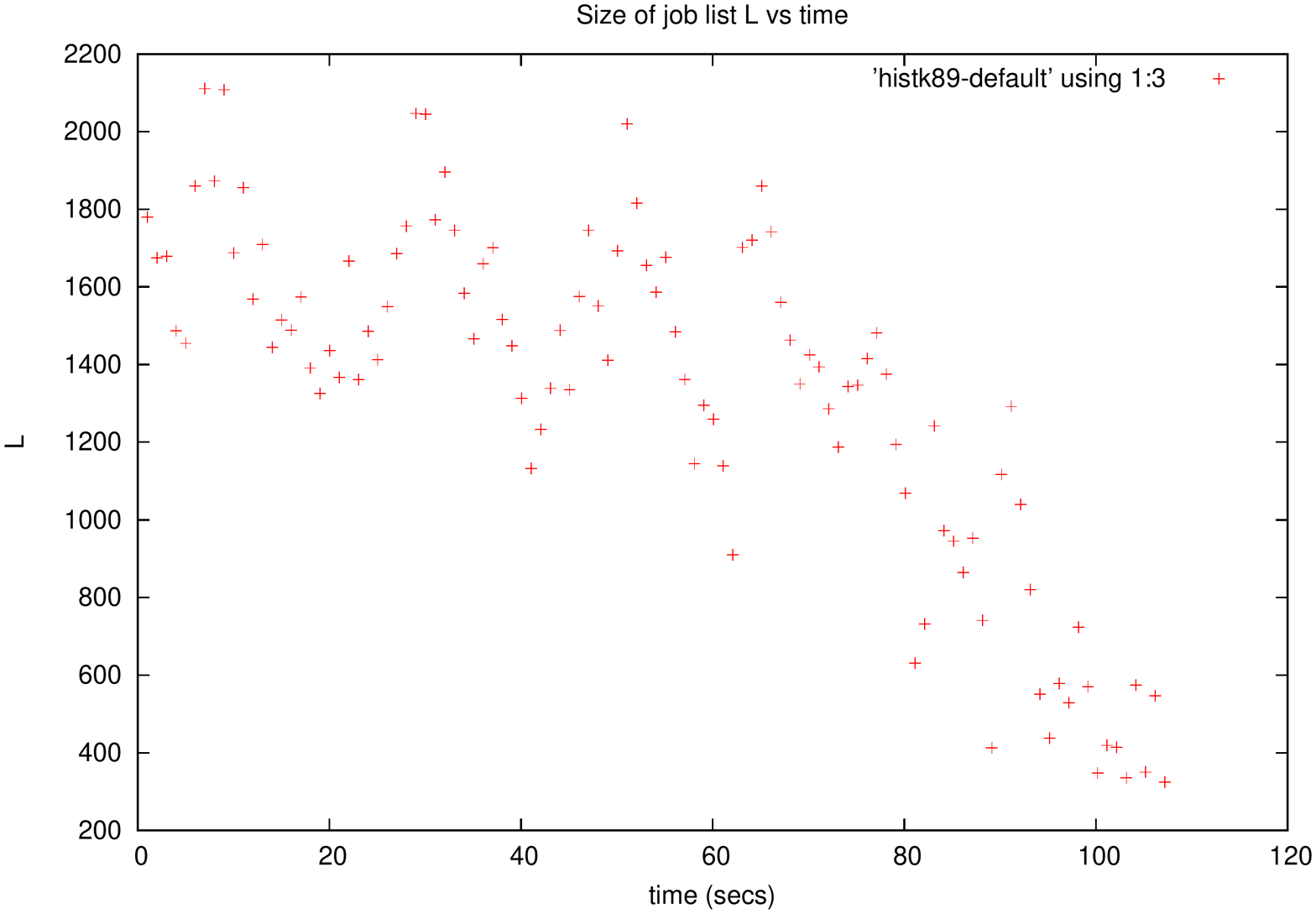}
\end{minipage}
\caption{Histograms for \mtop on $K_{8,9}$: busy workers (left) job list size (right)}
\label{fig:defplot}
\end{figure}

The amount of work contained in a subproblem can vary dramatically.
\mts can produce histograms to help understand and tune
its performance. We discuss three of these here: processor usage, job list size
and distribution of subproblem sizes.
Figure~\ref{fig:defplot}
shows the first two histograms for the \mtop run on \keightnine with default 
parameters (\ref{defpar}). 

We see the master struggling to keep 
workers busy despite having jobs available.  This suggests that we can 
improve performance with better parameters.
Here, a larger \emph{-scale} or \emph{-maxnodes} value may help,
since it will allow workers to do more work (assuming a sufficiently large 
subproblem) before contacting the master.

\begin{figure}[h!tbp]
\centering
\begin{minipage}{0.49\textwidth}
 \includegraphics[width=\textwidth]{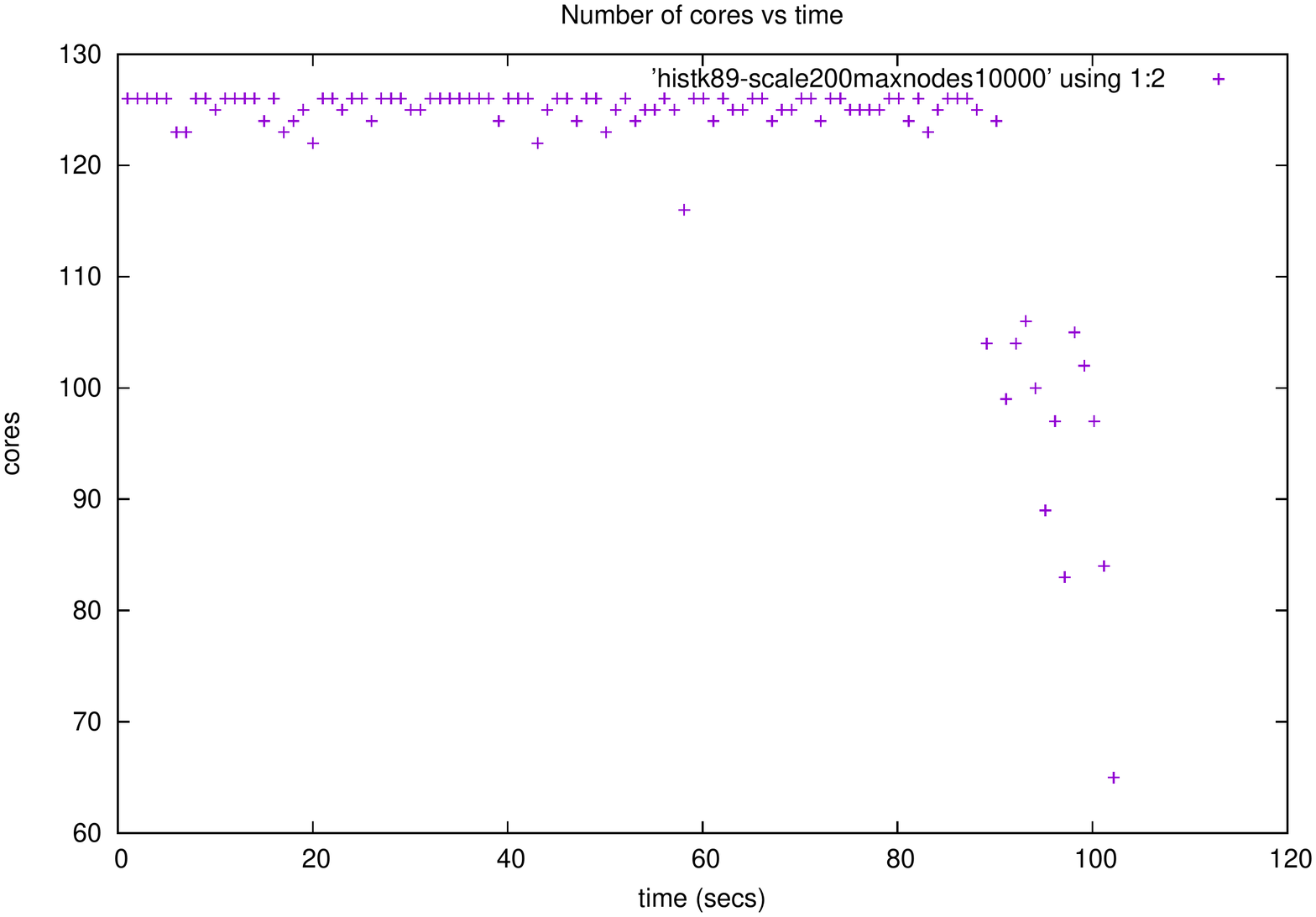}
\end{minipage}
\begin{minipage}{0.49\textwidth}
 \includegraphics[width=\textwidth]{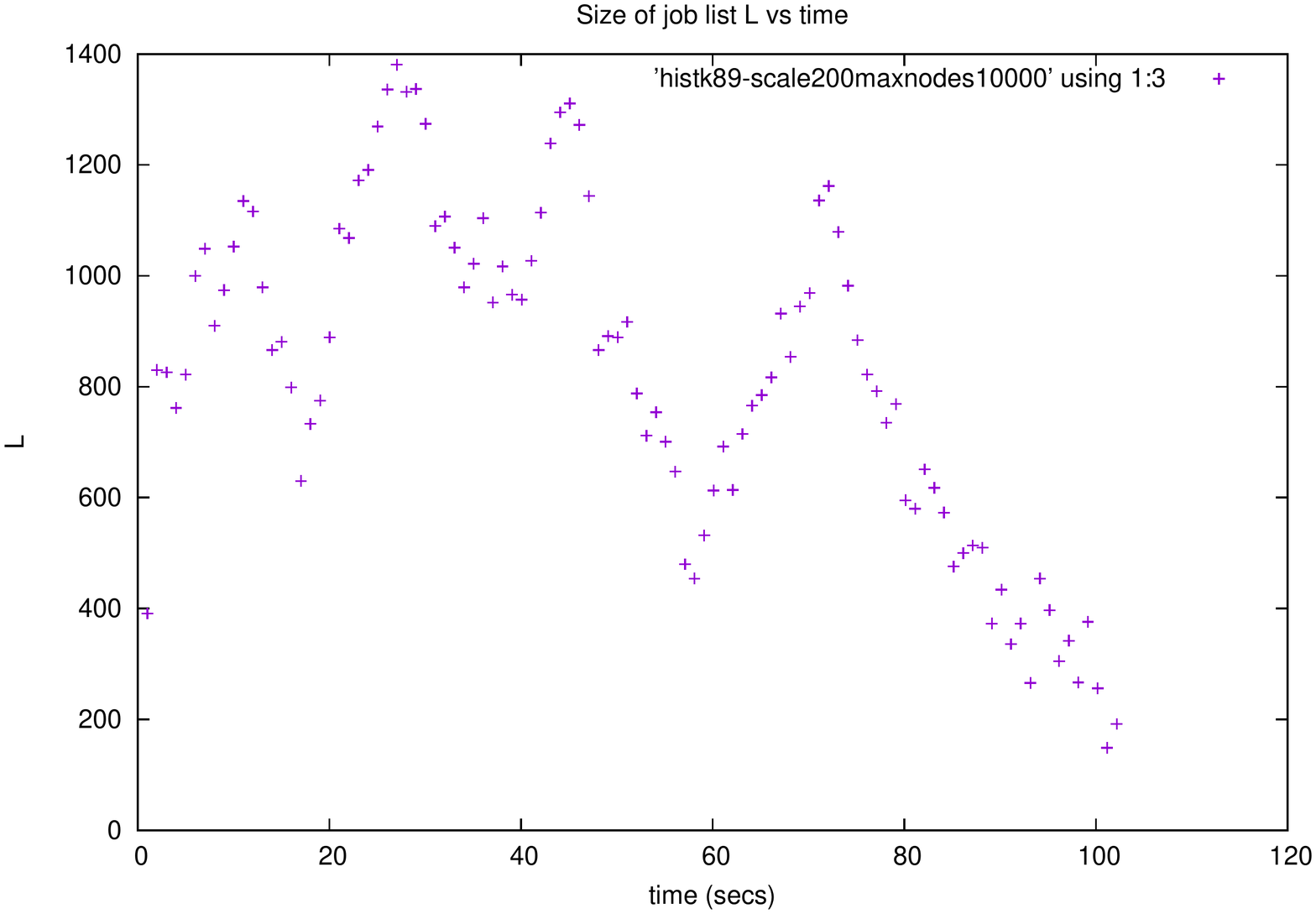}
\end{minipage}
\caption{Histograms with \emph{-scale} $200$ \emph{-maxnodes}
$10000$ on $K_{8,9}$: 
busy workers (l), joblist size (r)}
\label{fig:scaleplot}
\end{figure}

Figure~\ref{fig:scaleplot} shows the result of using $200$ for
\emph{-scale} and $10000$ for \emph{-maxnodes}.  These parameters produce less
than half the number of total number of jobs compared to the default
parameters, and increase overall performance by about five
percent on this input. 

\begin{figure}[h!tbp]
 \begin{minipage}{0.49\textwidth}
  \includegraphics[width=\textwidth]{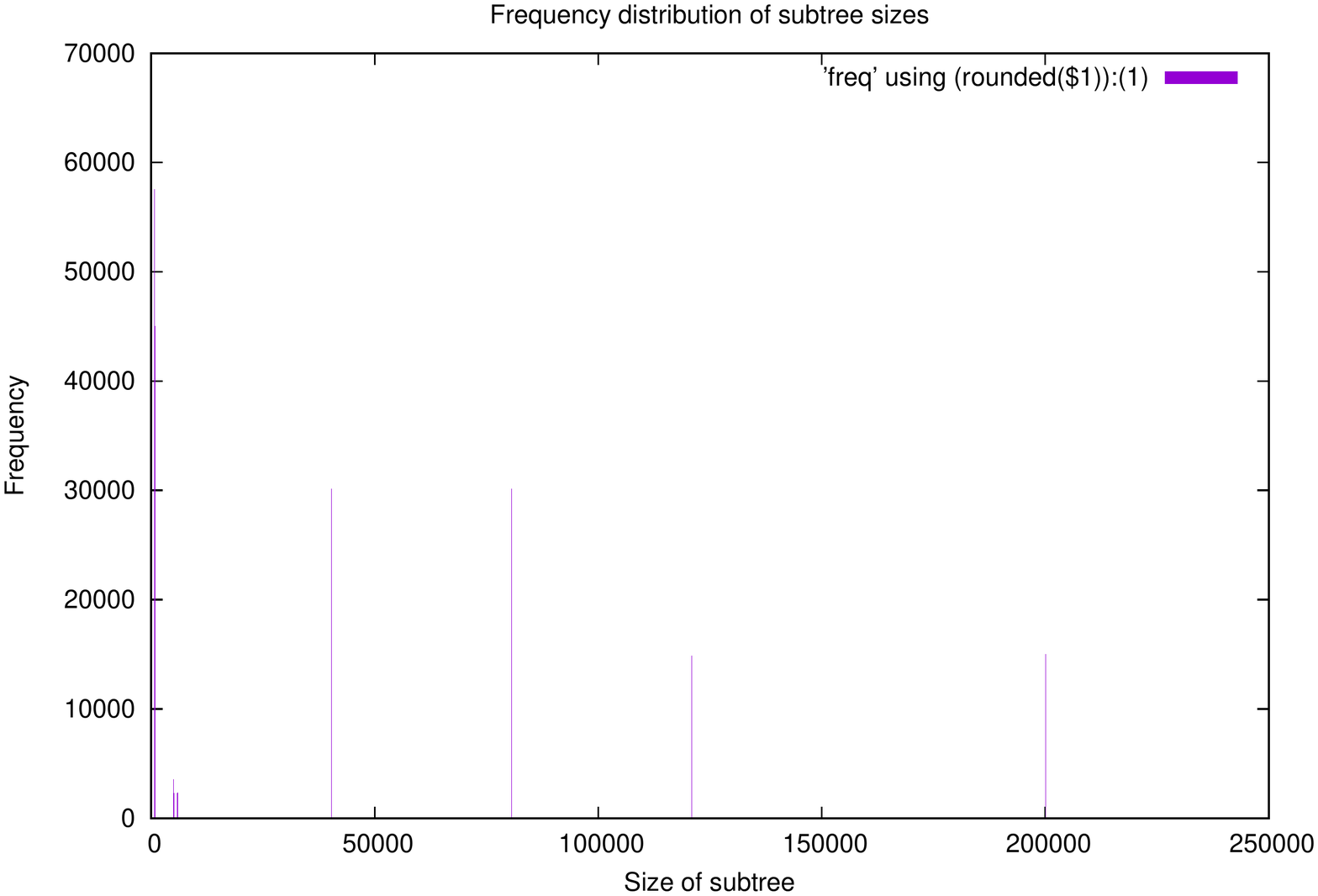}
 \end{minipage}
 \begin{minipage}{0.49\textwidth}
  \includegraphics[width=\textwidth]{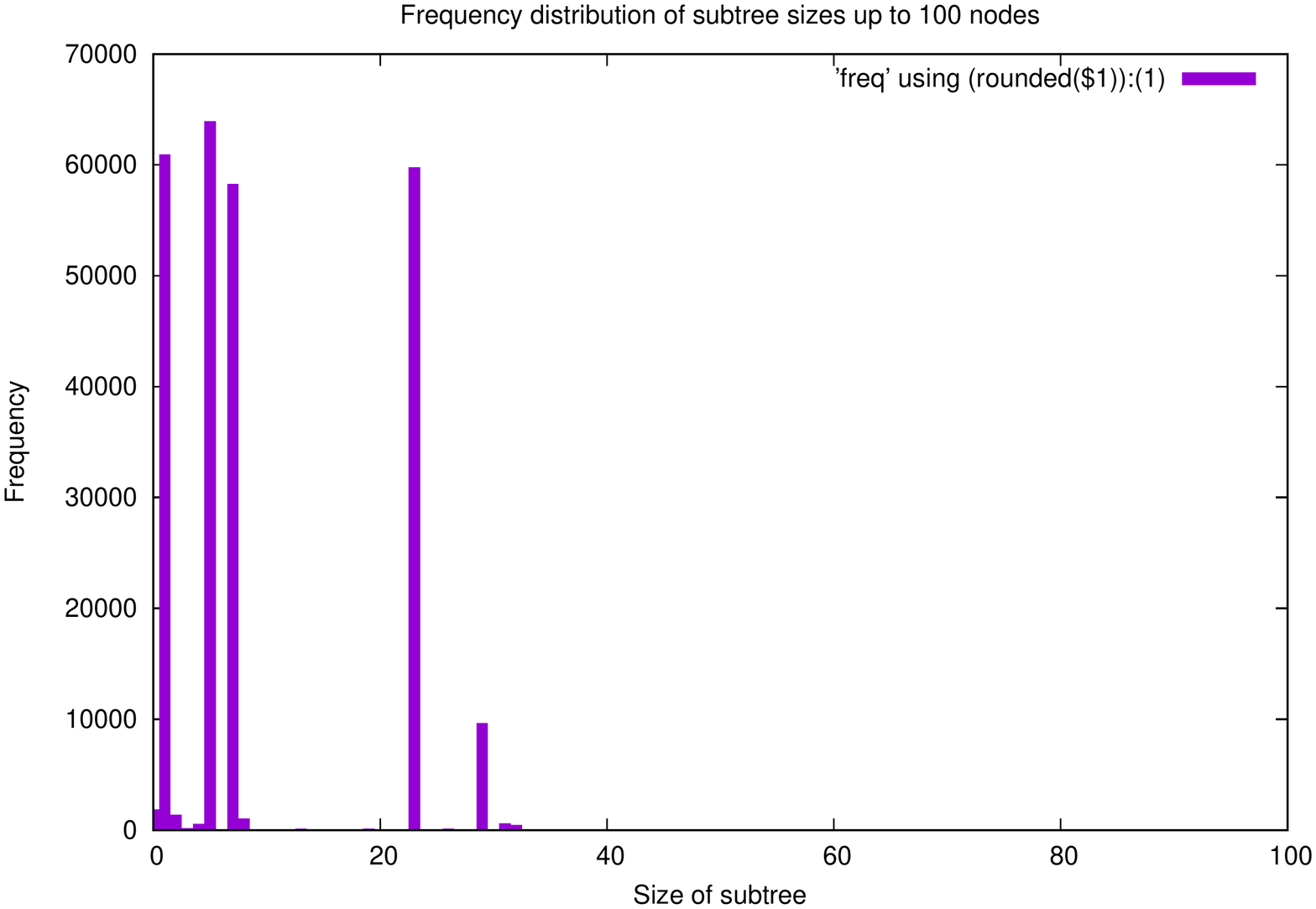}
 \end{minipage}
 \caption{Subproblem sizes for $K_{8,9}$: 
     all (left) small subproblems only (right)}
 \label{fig:freqplot}
\end{figure}

In addition to the performance histograms, \mts can generate
a \emph{frequency} file containing a list of values returned 
by each worker on the completion of each job. 
For the enumeration applications this is
normally the number of nodes visited by the worker during the job. 
Such a list
provides statistical information about the tree that is helpful
when tuning the parameters for better performance.  For example,
it may be helpful to implement and use pruning if many
jobs correspond to leaves.  Likewise, increasing the budget will
have limited effect if only few jobs use the full budget.
Figure~\ref{fig:freqplot} shows the distribution of subproblem
sizes that was produced in a run of \mtop on \keightnine with
default parameters (\ref{defpar}).
$L$ is usually large so the scaled budget constraint
of 200000 is normally in use. The left figure shows this constraint
was invoked about 15000 times. The right figure shows that most
subproblems have less than 40 nodes and so are not broken up.
The three spikes in the middle of the left figure are interesting and
show there are large numbers of subtrees with these specific sizes.
This is probably due to the high symmetry of the graph \keightnine.

\section{Conclusions}
\label{sec:concl}
We have presented a generic framework for parallelizing certain tree search codes requiring only
minimal changes to the legacy codes. Two features of our approach are that the parallelizing wrapper
does not need to be user modified and the modified legacy code can be tested in standalone
single processor mode. There is no separate library to install and just a few
routines need to be inserted in a user's existing library. Applying \mts to
four reverse search codes we obtained comparable results
to that previously obtained by the customized \mplrs wrapper applied to the
the \lrs code~\cite{AJ18a}. We expect that many other 
reverse search applications and  will obtain similar speedups when parallelized with \mts.

The application to SAT demonstrates the use of shared data, and the ease
with which a widely-used existing legacy code can be parallelized using \mts.  While \mtsat remains
work in progress, it shows some promise and further experimentation can likely improve performance.
Other ongoing work involves using \mts to parallelize existing integer programming solvers
that use the branch-and-bound approach.

\subparagraph*{Acknowledgements.}

This work was partially supported by JSPS
Kakenhi Grants 16H02785, 18K18027, 23700019 and 15H00847, Grant-in-Aid for Scientific Research
on Innovative Areas, `Exploring the Limits of Computation (ELC)'.

\bibliographystyle{spmpsci}
\bibliography{tutorial2}
\clearpage
\appendix
\section*{Appendix}

\begin{algorithm}[h!b]
 \caption{Master process}
 \label{alg:mts_master}
 \begin{algorithmic}[1]
  \Procedure{master}{\inputdata,
   \mymaxdepth, \maxnodes, \lmin, \lmax, \myscale, \numworkers}
   \State {\bf Send} (\inputdata) to each worker
   \State {\bf Create empty table} \sdata
   \State {\bf Create empty list} $L$
   \State {\bf Get} \startvertex from application, add to $L$
   \State $\mysize \gets \numworkers + 2$

   \While {$L$ is not empty or some worker is marked as working}
          \While {$L$ is not empty and some worker not marked as working}
             \If {$|L|<\mysize\cdot \lmin$} 
                \State $\maxd \gets \mymaxdepth$
             \Else \State $\maxd \gets \infty$
             \EndIf
             \If {$|L|>\mysize\cdot \lmax$} 
                \State $\nbudget \gets \myscale\cdot \maxnodes$
             \Else \State $\nbudget \gets \maxnodes$
             \EndIf
          \State {\bf Remove} next element \mystart from $L$
          \State {\bf Send} (\mystart, \maxd, \nbudget)
                            to first free worker $i$
	  \State {\bf Mark} $i$ as working
	  \State {\bf Send} any \shared in \sdata newer than $i$ has
          \EndWhile

          \For {each marked worker $i$}
            \State {\bf Check} for new message \unfinished from $i$
            \If {incoming message \unfinished from $i$}
               \State{\bf Join} list \unfinished to $L$
	       \State{\bf Receive} \shared update from $i$
	       \State{\bf Unmark} $i$ as working
	       \If {non-empty update}
		 \State{\bf Update} $i$'s \shared in \sdata
	       \EndIf
            \EndIf
          \EndFor

     \EndWhile
     \State {\bf Call} application with final set of \shared
     \State {\bf Send} \texttt{terminate} to all processes
  \EndProcedure
 \end{algorithmic}
\end{algorithm}
\begin{algorithm}[ht!]
 \caption{Worker process}
 \label{alg:mts_worker}
 \begin{algorithmic}[1]
  \Procedure{worker}{}
   \State{\bf Receive} (\inputdata) from master
   \State{\bf Create} empty \shared
   \While {\mytrue}
     \State{\bf Wait} for message from master
     \If {message is \texttt{terminate}}
       \State {\bf Exit}
     \EndIf
     \State{\bf Receive} (\startvertex, \mymaxdepth, \maxnodes)
     \State{\bf Receive} \shared updates, update local copy 
     \State{\bf Call} \search(\startvertex, \mymaxdepth, \maxnodes, \shared)
     \State{\bf Send} list of unfinished vertices to master
     \State{\bf Send} \shared update to master
     \State{\bf Send} output list to consumer
   \EndWhile
  \EndProcedure
 \end{algorithmic}
\end{algorithm}

\begin{algorithm}
 \caption{Consumer process}
 \label{alg:mts_consumer}
 \begin{algorithmic}[1]
  \Procedure{consumer}{}
   \While {\mytrue}
    \State{\bf Wait} for incoming message
    \If {message is \texttt{terminate}}
     \State {\bf Exit}
    \EndIf
    \State{\bf Output} this message
   \EndWhile
  \EndProcedure
 \end{algorithmic}
\end{algorithm}

\end{document}

%% file: plots/cactus-sat.tex
\begingroup
  \makeatletter
  \providecommand\color[2][]{%
    \GenericError{(gnuplot) \space\space\space\@spaces}{%
      Package color not loaded in conjunction with
      terminal option `colourtext'%
    }{See the gnuplot documentation for explanation.%
    }{Either use 'blacktext' in gnuplot or load the package
      color.sty in LaTeX.}%
    \renewcommand\color[2][]{}%
  }%
  \providecommand\includegraphics[2][]{%
    \GenericError{(gnuplot) \space\space\space\@spaces}{%
      Package graphicx or graphics not loaded%
    }{See the gnuplot documentation for explanation.%
    }{The gnuplot epslatex terminal needs graphicx.sty or graphics.sty.}%
    \renewcommand\includegraphics[2][]{}%
  }%
  \providecommand\rotatebox[2]{#2}%
  \@ifundefined{ifGPcolor}{%
    \newif\ifGPcolor
    \GPcolortrue
  }{}%
  \@ifundefined{ifGPblacktext}{%
    \newif\ifGPblacktext
    \GPblacktexttrue
  }{}%
  \let\gplgaddtomacro\g@addto@macro
  \gdef\gplbacktext{}%
  \gdef\gplfronttext{}%
  \makeatother
  \ifGPblacktext
    \def\colorrgb#1{}%
    \def\colorgray#1{}%
  \else
    \ifGPcolor
      \def\colorrgb#1{\color[rgb]{#1}}%
      \def\colorgray#1{\color[gray]{#1}}%
      \expandafter\def\csname LTw\endcsname{\color{white}}%
      \expandafter\def\csname LTb\endcsname{\color{black}}%
      \expandafter\def\csname LTa\endcsname{\color{black}}%
      \expandafter\def\csname LT0\endcsname{\color[rgb]{1,0,0}}%
      \expandafter\def\csname LT1\endcsname{\color[rgb]{0,1,0}}%
      \expandafter\def\csname LT2\endcsname{\color[rgb]{0,0,1}}%
      \expandafter\def\csname LT3\endcsname{\color[rgb]{1,0,1}}%
      \expandafter\def\csname LT4\endcsname{\color[rgb]{0,1,1}}%
      \expandafter\def\csname LT5\endcsname{\color[rgb]{1,1,0}}%
      \expandafter\def\csname LT6\endcsname{\color[rgb]{0,0,0}}%
      \expandafter\def\csname LT7\endcsname{\color[rgb]{1,0.3,0}}%
      \expandafter\def\csname LT8\endcsname{\color[rgb]{0.5,0.5,0.5}}%
    \else
      \def\colorrgb#1{\color{black}}%
      \def\colorgray#1{\color[gray]{#1}}%
      \expandafter\def\csname LTw\endcsname{\color{white}}%
      \expandafter\def\csname LTb\endcsname{\color{black}}%
      \expandafter\def\csname LTa\endcsname{\color{black}}%
      \expandafter\def\csname LT0\endcsname{\color{black}}%
      \expandafter\def\csname LT1\endcsname{\color{black}}%
      \expandafter\def\csname LT2\endcsname{\color{black}}%
      \expandafter\def\csname LT3\endcsname{\color{black}}%
      \expandafter\def\csname LT4\endcsname{\color{black}}%
      \expandafter\def\csname LT5\endcsname{\color{black}}%
      \expandafter\def\csname LT6\endcsname{\color{black}}%
      \expandafter\def\csname LT7\endcsname{\color{black}}%
      \expandafter\def\csname LT8\endcsname{\color{black}}%
    \fi
  \fi
    \setlength{\unitlength}{0.0500bp}%
    \ifx\gptboxheight\undefined%
      \newlength{\gptboxheight}%
      \newlength{\gptboxwidth}%
      \newsavebox{\gptboxtext}%
    \fi%
    \setlength{\fboxrule}{0.5pt}%
    \setlength{\fboxsep}{1pt}%
\begin{picture}(5040.00,3600.00)%
    \gplgaddtomacro\gplbacktext{%
      \csname LTb\endcsname%
      \put(946,704){\makebox(0,0)[r]{\strut{}$0$}}%
      \put(946,1143){\makebox(0,0)[r]{\strut{}$200$}}%
      \put(946,1581){\makebox(0,0)[r]{\strut{}$400$}}%
      \put(946,2020){\makebox(0,0)[r]{\strut{}$600$}}%
      \put(946,2458){\makebox(0,0)[r]{\strut{}$800$}}%
      \put(946,2897){\makebox(0,0)[r]{\strut{}$1000$}}%
      \put(946,3335){\makebox(0,0)[r]{\strut{}$1200$}}%
      \put(1543,484){\makebox(0,0){\strut{}$10$}}%
      \put(2060,484){\makebox(0,0){\strut{}$20$}}%
      \put(2576,484){\makebox(0,0){\strut{}$30$}}%
      \put(3093,484){\makebox(0,0){\strut{}$40$}}%
      \put(3610,484){\makebox(0,0){\strut{}$50$}}%
      \put(4126,484){\makebox(0,0){\strut{}$60$}}%
      \put(4643,484){\makebox(0,0){\strut{}$70$}}%
    }%
    \gplgaddtomacro\gplfronttext{%
      \csname LTb\endcsname%
      \put(176,2019){\rotatebox{-270}{\makebox(0,0){\strut{}time (s)}}}%
      \put(2860,154){\makebox(0,0){\strut{}problems solved (out of 100)}}%
      \csname LTb\endcsname%
      \put(3656,1977){\makebox(0,0)[r]{\strut{}\minisat}}%
      \csname LTb\endcsname%
      \put(3656,1757){\makebox(0,0)[r]{\strut{}\mtsat (16)}}%
      \csname LTb\endcsname%
      \put(3656,1537){\makebox(0,0)[r]{\strut{}\mtsat (32)}}%
      \csname LTb\endcsname%
      \put(3656,1317){\makebox(0,0)[r]{\strut{}\mtsat (64)}}%
      \csname LTb\endcsname%
      \put(3656,1097){\makebox(0,0)[r]{\strut{}\mtsat (128)}}%
      \csname LTb\endcsname%
      \put(3656,877){\makebox(0,0)[r]{\strut{}\mtsat (192)}}%
    }%
    \gplbacktext
    \put(0,0){\includegraphics{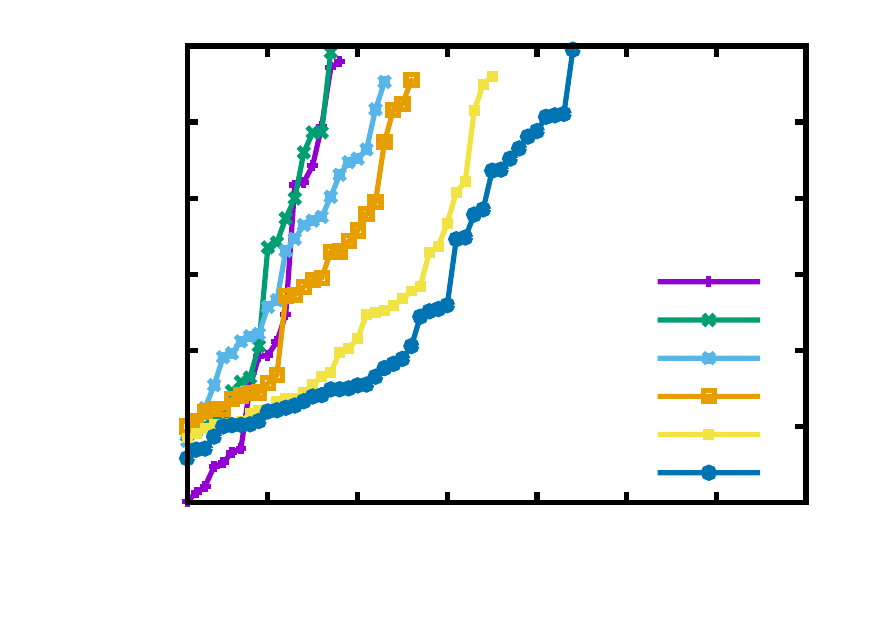}}%
    \gplfronttext
  \end{picture}%
\endgroup

%% file: plots/cactus-sat-confbudg.tex
\begingroup
  \makeatletter
  \providecommand\color[2][]{%
    \GenericError{(gnuplot) \space\space\space\@spaces}{%
      Package color not loaded in conjunction with
      terminal option `colourtext'%
    }{See the gnuplot documentation for explanation.%
    }{Either use 'blacktext' in gnuplot or load the package
      color.sty in LaTeX.}%
    \renewcommand\color[2][]{}%
  }%
  \providecommand\includegraphics[2][]{%
    \GenericError{(gnuplot) \space\space\space\@spaces}{%
      Package graphicx or graphics not loaded%
    }{See the gnuplot documentation for explanation.%
    }{The gnuplot epslatex terminal needs graphicx.sty or graphics.sty.}%
    \renewcommand\includegraphics[2][]{}%
  }%
  \providecommand\rotatebox[2]{#2}%
  \@ifundefined{ifGPcolor}{%
    \newif\ifGPcolor
    \GPcolortrue
  }{}%
  \@ifundefined{ifGPblacktext}{%
    \newif\ifGPblacktext
    \GPblacktexttrue
  }{}%
  \let\gplgaddtomacro\g@addto@macro
  \gdef\gplbacktext{}%
  \gdef\gplfronttext{}%
  \makeatother
  \ifGPblacktext
    \def\colorrgb#1{}%
    \def\colorgray#1{}%
  \else
    \ifGPcolor
      \def\colorrgb#1{\color[rgb]{#1}}%
      \def\colorgray#1{\color[gray]{#1}}%
      \expandafter\def\csname LTw\endcsname{\color{white}}%
      \expandafter\def\csname LTb\endcsname{\color{black}}%
      \expandafter\def\csname LTa\endcsname{\color{black}}%
      \expandafter\def\csname LT0\endcsname{\color[rgb]{1,0,0}}%
      \expandafter\def\csname LT1\endcsname{\color[rgb]{0,1,0}}%
      \expandafter\def\csname LT2\endcsname{\color[rgb]{0,0,1}}%
      \expandafter\def\csname LT3\endcsname{\color[rgb]{1,0,1}}%
      \expandafter\def\csname LT4\endcsname{\color[rgb]{0,1,1}}%
      \expandafter\def\csname LT5\endcsname{\color[rgb]{1,1,0}}%
      \expandafter\def\csname LT6\endcsname{\color[rgb]{0,0,0}}%
      \expandafter\def\csname LT7\endcsname{\color[rgb]{1,0.3,0}}%
      \expandafter\def\csname LT8\endcsname{\color[rgb]{0.5,0.5,0.5}}%
    \else
      \def\colorrgb#1{\color{black}}%
      \def\colorgray#1{\color[gray]{#1}}%
      \expandafter\def\csname LTw\endcsname{\color{white}}%
      \expandafter\def\csname LTb\endcsname{\color{black}}%
      \expandafter\def\csname LTa\endcsname{\color{black}}%
      \expandafter\def\csname LT0\endcsname{\color{black}}%
      \expandafter\def\csname LT1\endcsname{\color{black}}%
      \expandafter\def\csname LT2\endcsname{\color{black}}%
      \expandafter\def\csname LT3\endcsname{\color{black}}%
      \expandafter\def\csname LT4\endcsname{\color{black}}%
      \expandafter\def\csname LT5\endcsname{\color{black}}%
      \expandafter\def\csname LT6\endcsname{\color{black}}%
      \expandafter\def\csname LT7\endcsname{\color{black}}%
      \expandafter\def\csname LT8\endcsname{\color{black}}%
    \fi
  \fi
    \setlength{\unitlength}{0.0500bp}%
    \ifx\gptboxheight\undefined%
      \newlength{\gptboxheight}%
      \newlength{\gptboxwidth}%
      \newsavebox{\gptboxtext}%
    \fi%
    \setlength{\fboxrule}{0.5pt}%
    \setlength{\fboxsep}{1pt}%
\begin{picture}(5040.00,3600.00)%
    \gplgaddtomacro\gplbacktext{%
      \csname LTb\endcsname%
      \put(946,704){\makebox(0,0)[r]{\strut{}$0$}}%
      \put(946,1143){\makebox(0,0)[r]{\strut{}$200$}}%
      \put(946,1581){\makebox(0,0)[r]{\strut{}$400$}}%
      \put(946,2020){\makebox(0,0)[r]{\strut{}$600$}}%
      \put(946,2458){\makebox(0,0)[r]{\strut{}$800$}}%
      \put(946,2897){\makebox(0,0)[r]{\strut{}$1000$}}%
      \put(946,3335){\makebox(0,0)[r]{\strut{}$1200$}}%
      \put(1543,484){\makebox(0,0){\strut{}$10$}}%
      \put(2060,484){\makebox(0,0){\strut{}$20$}}%
      \put(2576,484){\makebox(0,0){\strut{}$30$}}%
      \put(3093,484){\makebox(0,0){\strut{}$40$}}%
      \put(3610,484){\makebox(0,0){\strut{}$50$}}%
      \put(4126,484){\makebox(0,0){\strut{}$60$}}%
      \put(4643,484){\makebox(0,0){\strut{}$70$}}%
    }%
    \gplgaddtomacro\gplfronttext{%
      \csname LTb\endcsname%
      \put(176,2019){\rotatebox{-270}{\makebox(0,0){\strut{}time (s)}}}%
      \put(2860,154){\makebox(0,0){\strut{}problems solved (out of 100)}}%
      \csname LTb\endcsname%
      \put(3656,1977){\makebox(0,0)[r]{\strut{}\minisat}}%
      \csname LTb\endcsname%
      \put(3656,1757){\makebox(0,0)[r]{\strut{}\mtsat (16)}}%
      \csname LTb\endcsname%
      \put(3656,1537){\makebox(0,0)[r]{\strut{}\mtsat (32)}}%
      \csname LTb\endcsname%
      \put(3656,1317){\makebox(0,0)[r]{\strut{}\mtsat (64)}}%
      \csname LTb\endcsname%
      \put(3656,1097){\makebox(0,0)[r]{\strut{}\mtsat (128)}}%
      \csname LTb\endcsname%
      \put(3656,877){\makebox(0,0)[r]{\strut{}\mtsat (192)}}%
    }%
    \gplbacktext
    \put(0,0){\includegraphics{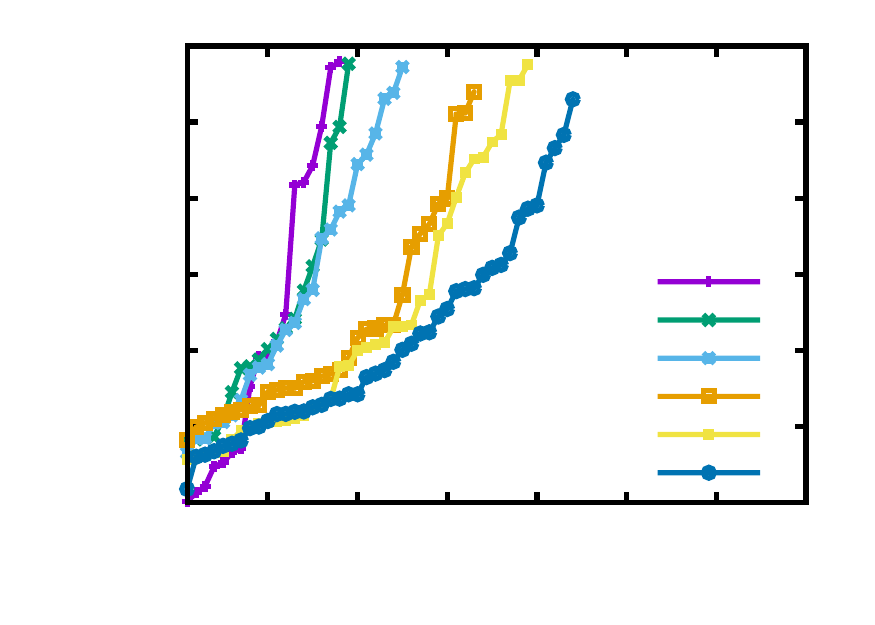}}%
    \gplfronttext
  \end{picture}%
\endgroup

%% file: plots/cactus-sat-glucose.tex
\begingroup
  \makeatletter
  \providecommand\color[2][]{%
    \GenericError{(gnuplot) \space\space\space\@spaces}{%
      Package color not loaded in conjunction with
      terminal option `colourtext'%
    }{See the gnuplot documentation for explanation.%
    }{Either use 'blacktext' in gnuplot or load the package
      color.sty in LaTeX.}%
    \renewcommand\color[2][]{}%
  }%
  \providecommand\includegraphics[2][]{%
    \GenericError{(gnuplot) \space\space\space\@spaces}{%
      Package graphicx or graphics not loaded%
    }{See the gnuplot documentation for explanation.%
    }{The gnuplot epslatex terminal needs graphicx.sty or graphics.sty.}%
    \renewcommand\includegraphics[2][]{}%
  }%
  \providecommand\rotatebox[2]{#2}%
  \@ifundefined{ifGPcolor}{%
    \newif\ifGPcolor
    \GPcolortrue
  }{}%
  \@ifundefined{ifGPblacktext}{%
    \newif\ifGPblacktext
    \GPblacktexttrue
  }{}%
  \let\gplgaddtomacro\g@addto@macro
  \gdef\gplbacktext{}%
  \gdef\gplfronttext{}%
  \makeatother
  \ifGPblacktext
    \def\colorrgb#1{}%
    \def\colorgray#1{}%
  \else
    \ifGPcolor
      \def\colorrgb#1{\color[rgb]{#1}}%
      \def\colorgray#1{\color[gray]{#1}}%
      \expandafter\def\csname LTw\endcsname{\color{white}}%
      \expandafter\def\csname LTb\endcsname{\color{black}}%
      \expandafter\def\csname LTa\endcsname{\color{black}}%
      \expandafter\def\csname LT0\endcsname{\color[rgb]{1,0,0}}%
      \expandafter\def\csname LT1\endcsname{\color[rgb]{0,1,0}}%
      \expandafter\def\csname LT2\endcsname{\color[rgb]{0,0,1}}%
      \expandafter\def\csname LT3\endcsname{\color[rgb]{1,0,1}}%
      \expandafter\def\csname LT4\endcsname{\color[rgb]{0,1,1}}%
      \expandafter\def\csname LT5\endcsname{\color[rgb]{1,1,0}}%
      \expandafter\def\csname LT6\endcsname{\color[rgb]{0,0,0}}%
      \expandafter\def\csname LT7\endcsname{\color[rgb]{1,0.3,0}}%
      \expandafter\def\csname LT8\endcsname{\color[rgb]{0.5,0.5,0.5}}%
    \else
      \def\colorrgb#1{\color{black}}%
      \def\colorgray#1{\color[gray]{#1}}%
      \expandafter\def\csname LTw\endcsname{\color{white}}%
      \expandafter\def\csname LTb\endcsname{\color{black}}%
      \expandafter\def\csname LTa\endcsname{\color{black}}%
      \expandafter\def\csname LT0\endcsname{\color{black}}%
      \expandafter\def\csname LT1\endcsname{\color{black}}%
      \expandafter\def\csname LT2\endcsname{\color{black}}%
      \expandafter\def\csname LT3\endcsname{\color{black}}%
      \expandafter\def\csname LT4\endcsname{\color{black}}%
      \expandafter\def\csname LT5\endcsname{\color{black}}%
      \expandafter\def\csname LT6\endcsname{\color{black}}%
      \expandafter\def\csname LT7\endcsname{\color{black}}%
      \expandafter\def\csname LT8\endcsname{\color{black}}%
    \fi
  \fi
    \setlength{\unitlength}{0.0500bp}%
    \ifx\gptboxheight\undefined%
      \newlength{\gptboxheight}%
      \newlength{\gptboxwidth}%
      \newsavebox{\gptboxtext}%
    \fi%
    \setlength{\fboxrule}{0.5pt}%
    \setlength{\fboxsep}{1pt}%
\begin{picture}(5040.00,3600.00)%
    \gplgaddtomacro\gplbacktext{%
      \csname LTb\endcsname%
      \put(946,704){\makebox(0,0)[r]{\strut{}$0$}}%
      \put(946,1143){\makebox(0,0)[r]{\strut{}$200$}}%
      \put(946,1581){\makebox(0,0)[r]{\strut{}$400$}}%
      \put(946,2020){\makebox(0,0)[r]{\strut{}$600$}}%
      \put(946,2458){\makebox(0,0)[r]{\strut{}$800$}}%
      \put(946,2897){\makebox(0,0)[r]{\strut{}$1000$}}%
      \put(946,3335){\makebox(0,0)[r]{\strut{}$1200$}}%
      \put(1543,484){\makebox(0,0){\strut{}$10$}}%
      \put(2060,484){\makebox(0,0){\strut{}$20$}}%
      \put(2576,484){\makebox(0,0){\strut{}$30$}}%
      \put(3093,484){\makebox(0,0){\strut{}$40$}}%
      \put(3610,484){\makebox(0,0){\strut{}$50$}}%
      \put(4126,484){\makebox(0,0){\strut{}$60$}}%
      \put(4643,484){\makebox(0,0){\strut{}$70$}}%
    }%
    \gplgaddtomacro\gplfronttext{%
      \csname LTb\endcsname%
      \put(176,2019){\rotatebox{-270}{\makebox(0,0){\strut{}time (s)}}}%
      \put(2860,154){\makebox(0,0){\strut{}problems solved (out of 100)}}%
      \csname LTb\endcsname%
      \put(3656,1977){\makebox(0,0)[r]{\strut{}\glucose}}%
      \csname LTb\endcsname%
      \put(3656,1757){\makebox(0,0)[r]{\strut{}\mtsatglucose (16)}}%
      \csname LTb\endcsname%
      \put(3656,1537){\makebox(0,0)[r]{\strut{}\mtsatglucose (32)}}%
      \csname LTb\endcsname%
      \put(3656,1317){\makebox(0,0)[r]{\strut{}\mtsatglucose (64)}}%
      \csname LTb\endcsname%
      \put(3656,1097){\makebox(0,0)[r]{\strut{}\mtsatglucose (128)}}%
      \csname LTb\endcsname%
      \put(3656,877){\makebox(0,0)[r]{\strut{}\mtsatglucose (192)}}%
    }%
    \gplbacktext
    \put(0,0){\includegraphics{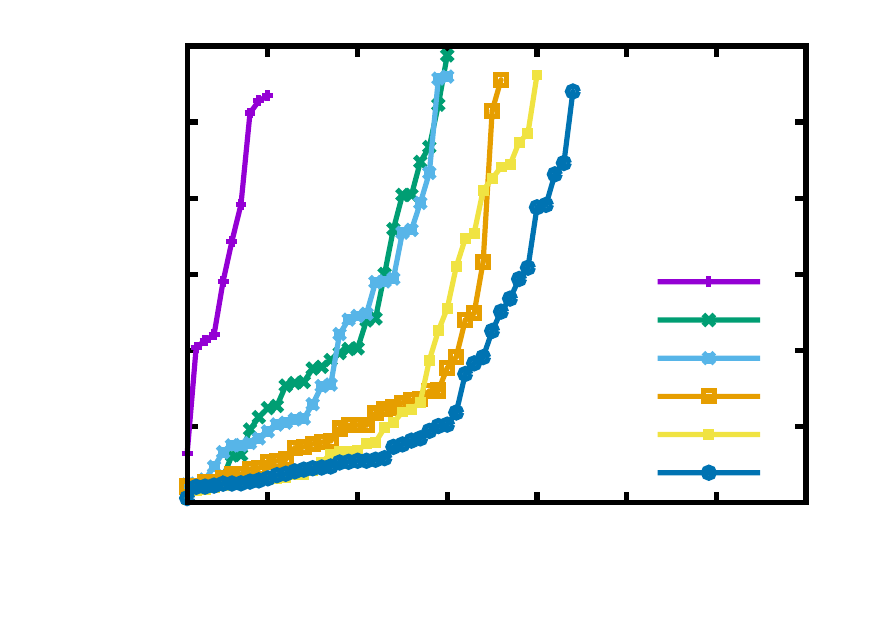}}%
    \gplfronttext
  \end{picture}%
\endgroup

%% file: plots/cactus-sat-glucose-confbudg.tex
\begingroup
  \makeatletter
  \providecommand\color[2][]{%
    \GenericError{(gnuplot) \space\space\space\@spaces}{%
      Package color not loaded in conjunction with
      terminal option `colourtext'%
    }{See the gnuplot documentation for explanation.%
    }{Either use 'blacktext' in gnuplot or load the package
      color.sty in LaTeX.}%
    \renewcommand\color[2][]{}%
  }%
  \providecommand\includegraphics[2][]{%
    \GenericError{(gnuplot) \space\space\space\@spaces}{%
      Package graphicx or graphics not loaded%
    }{See the gnuplot documentation for explanation.%
    }{The gnuplot epslatex terminal needs graphicx.sty or graphics.sty.}%
    \renewcommand\includegraphics[2][]{}%
  }%
  \providecommand\rotatebox[2]{#2}%
  \@ifundefined{ifGPcolor}{%
    \newif\ifGPcolor
    \GPcolortrue
  }{}%
  \@ifundefined{ifGPblacktext}{%
    \newif\ifGPblacktext
    \GPblacktexttrue
  }{}%
  \let\gplgaddtomacro\g@addto@macro
  \gdef\gplbacktext{}%
  \gdef\gplfronttext{}%
  \makeatother
  \ifGPblacktext
    \def\colorrgb#1{}%
    \def\colorgray#1{}%
  \else
    \ifGPcolor
      \def\colorrgb#1{\color[rgb]{#1}}%
      \def\colorgray#1{\color[gray]{#1}}%
      \expandafter\def\csname LTw\endcsname{\color{white}}%
      \expandafter\def\csname LTb\endcsname{\color{black}}%
      \expandafter\def\csname LTa\endcsname{\color{black}}%
      \expandafter\def\csname LT0\endcsname{\color[rgb]{1,0,0}}%
      \expandafter\def\csname LT1\endcsname{\color[rgb]{0,1,0}}%
      \expandafter\def\csname LT2\endcsname{\color[rgb]{0,0,1}}%
      \expandafter\def\csname LT3\endcsname{\color[rgb]{1,0,1}}%
      \expandafter\def\csname LT4\endcsname{\color[rgb]{0,1,1}}%
      \expandafter\def\csname LT5\endcsname{\color[rgb]{1,1,0}}%
      \expandafter\def\csname LT6\endcsname{\color[rgb]{0,0,0}}%
      \expandafter\def\csname LT7\endcsname{\color[rgb]{1,0.3,0}}%
      \expandafter\def\csname LT8\endcsname{\color[rgb]{0.5,0.5,0.5}}%
    \else
      \def\colorrgb#1{\color{black}}%
      \def\colorgray#1{\color[gray]{#1}}%
      \expandafter\def\csname LTw\endcsname{\color{white}}%
      \expandafter\def\csname LTb\endcsname{\color{black}}%
      \expandafter\def\csname LTa\endcsname{\color{black}}%
      \expandafter\def\csname LT0\endcsname{\color{black}}%
      \expandafter\def\csname LT1\endcsname{\color{black}}%
      \expandafter\def\csname LT2\endcsname{\color{black}}%
      \expandafter\def\csname LT3\endcsname{\color{black}}%
      \expandafter\def\csname LT4\endcsname{\color{black}}%
      \expandafter\def\csname LT5\endcsname{\color{black}}%
      \expandafter\def\csname LT6\endcsname{\color{black}}%
      \expandafter\def\csname LT7\endcsname{\color{black}}%
      \expandafter\def\csname LT8\endcsname{\color{black}}%
    \fi
  \fi
    \setlength{\unitlength}{0.0500bp}%
    \ifx\gptboxheight\undefined%
      \newlength{\gptboxheight}%
      \newlength{\gptboxwidth}%
      \newsavebox{\gptboxtext}%
    \fi%
    \setlength{\fboxrule}{0.5pt}%
    \setlength{\fboxsep}{1pt}%
\begin{picture}(5040.00,3600.00)%
    \gplgaddtomacro\gplbacktext{%
      \csname LTb\endcsname%
      \put(946,704){\makebox(0,0)[r]{\strut{}$0$}}%
      \put(946,1143){\makebox(0,0)[r]{\strut{}$200$}}%
      \put(946,1581){\makebox(0,0)[r]{\strut{}$400$}}%
      \put(946,2020){\makebox(0,0)[r]{\strut{}$600$}}%
      \put(946,2458){\makebox(0,0)[r]{\strut{}$800$}}%
      \put(946,2897){\makebox(0,0)[r]{\strut{}$1000$}}%
      \put(946,3335){\makebox(0,0)[r]{\strut{}$1200$}}%
      \put(1543,484){\makebox(0,0){\strut{}$10$}}%
      \put(2060,484){\makebox(0,0){\strut{}$20$}}%
      \put(2576,484){\makebox(0,0){\strut{}$30$}}%
      \put(3093,484){\makebox(0,0){\strut{}$40$}}%
      \put(3610,484){\makebox(0,0){\strut{}$50$}}%
      \put(4126,484){\makebox(0,0){\strut{}$60$}}%
      \put(4643,484){\makebox(0,0){\strut{}$70$}}%
    }%
    \gplgaddtomacro\gplfronttext{%
      \csname LTb\endcsname%
      \put(176,2019){\rotatebox{-270}{\makebox(0,0){\strut{}time (s)}}}%
      \put(2860,154){\makebox(0,0){\strut{}problems solved (out of 100)}}%
      \csname LTb\endcsname%
      \put(3656,1977){\makebox(0,0)[r]{\strut{}\glucose}}%
      \csname LTb\endcsname%
      \put(3656,1757){\makebox(0,0)[r]{\strut{}\mtsatglucose (16)}}%
      \csname LTb\endcsname%
      \put(3656,1537){\makebox(0,0)[r]{\strut{}\mtsatglucose (32)}}%
      \csname LTb\endcsname%
      \put(3656,1317){\makebox(0,0)[r]{\strut{}\mtsatglucose (64)}}%
      \csname LTb\endcsname%
      \put(3656,1097){\makebox(0,0)[r]{\strut{}\mtsatglucose (128)}}%
      \csname LTb\endcsname%
      \put(3656,877){\makebox(0,0)[r]{\strut{}\mtsatglucose (192)}}%
    }%
    \gplbacktext
    \put(0,0){\includegraphics{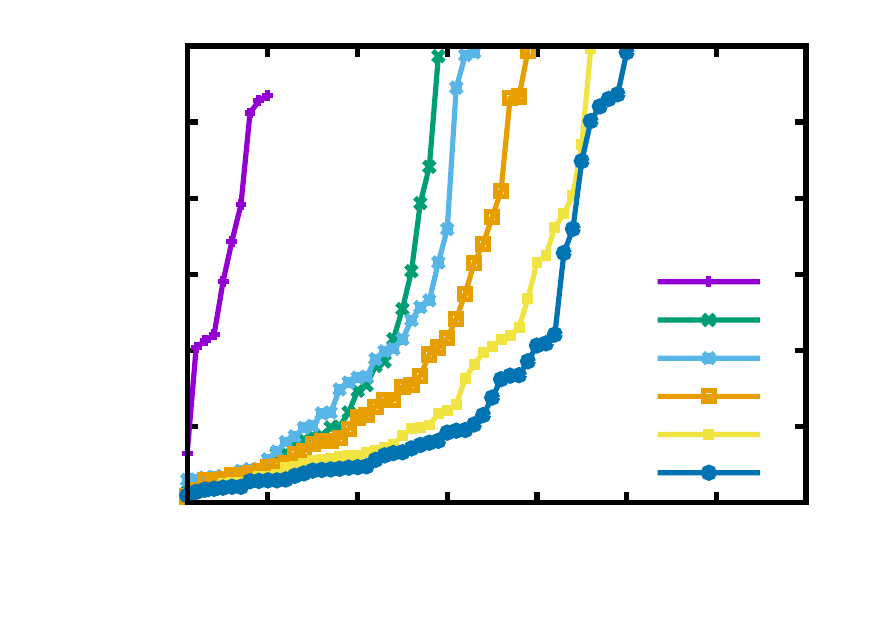}}%
    \gplfronttext
  \end{picture}%
\endgroup

%% file: plots/efficiency-mtopsorts.tex
\begingroup
  \makeatletter
  \providecommand\color[2][]{%
    \GenericError{(gnuplot) \space\space\space\@spaces}{%
      Package color not loaded in conjunction with
      terminal option `colourtext'%
    }{See the gnuplot documentation for explanation.%
    }{Either use 'blacktext' in gnuplot or load the package
      color.sty in LaTeX.}%
    \renewcommand\color[2][]{}%
  }%
  \providecommand\includegraphics[2][]{%
    \GenericError{(gnuplot) \space\space\space\@spaces}{%
      Package graphicx or graphics not loaded%
    }{See the gnuplot documentation for explanation.%
    }{The gnuplot epslatex terminal needs graphicx.sty or graphics.sty.}%
    \renewcommand\includegraphics[2][]{}%
  }%
  \providecommand\rotatebox[2]{#2}%
  \@ifundefined{ifGPcolor}{%
    \newif\ifGPcolor
    \GPcolortrue
  }{}%
  \@ifundefined{ifGPblacktext}{%
    \newif\ifGPblacktext
    \GPblacktexttrue
  }{}%
  \let\gplgaddtomacro\g@addto@macro
  \gdef\gplbacktext{}%
  \gdef\gplfronttext{}%
  \makeatother
  \ifGPblacktext
    \def\colorrgb#1{}%
    \def\colorgray#1{}%
  \else
    \ifGPcolor
      \def\colorrgb#1{\color[rgb]{#1}}%
      \def\colorgray#1{\color[gray]{#1}}%
      \expandafter\def\csname LTw\endcsname{\color{white}}%
      \expandafter\def\csname LTb\endcsname{\color{black}}%
      \expandafter\def\csname LTa\endcsname{\color{black}}%
      \expandafter\def\csname LT0\endcsname{\color[rgb]{1,0,0}}%
      \expandafter\def\csname LT1\endcsname{\color[rgb]{0,1,0}}%
      \expandafter\def\csname LT2\endcsname{\color[rgb]{0,0,1}}%
      \expandafter\def\csname LT3\endcsname{\color[rgb]{1,0,1}}%
      \expandafter\def\csname LT4\endcsname{\color[rgb]{0,1,1}}%
      \expandafter\def\csname LT5\endcsname{\color[rgb]{1,1,0}}%
      \expandafter\def\csname LT6\endcsname{\color[rgb]{0,0,0}}%
      \expandafter\def\csname LT7\endcsname{\color[rgb]{1,0.3,0}}%
      \expandafter\def\csname LT8\endcsname{\color[rgb]{0.5,0.5,0.5}}%
    \else
      \def\colorrgb#1{\color{black}}%
      \def\colorgray#1{\color[gray]{#1}}%
      \expandafter\def\csname LTw\endcsname{\color{white}}%
      \expandafter\def\csname LTb\endcsname{\color{black}}%
      \expandafter\def\csname LTa\endcsname{\color{black}}%
      \expandafter\def\csname LT0\endcsname{\color{black}}%
      \expandafter\def\csname LT1\endcsname{\color{black}}%
      \expandafter\def\csname LT2\endcsname{\color{black}}%
      \expandafter\def\csname LT3\endcsname{\color{black}}%
      \expandafter\def\csname LT4\endcsname{\color{black}}%
      \expandafter\def\csname LT5\endcsname{\color{black}}%
      \expandafter\def\csname LT6\endcsname{\color{black}}%
      \expandafter\def\csname LT7\endcsname{\color{black}}%
      \expandafter\def\csname LT8\endcsname{\color{black}}%
    \fi
  \fi
    \setlength{\unitlength}{0.0500bp}%
    \ifx\gptboxheight\undefined%
      \newlength{\gptboxheight}%
      \newlength{\gptboxwidth}%
      \newsavebox{\gptboxtext}%
    \fi%
    \setlength{\fboxrule}{0.5pt}%
    \setlength{\fboxsep}{1pt}%
\begin{picture}(5040.00,3600.00)%
    \gplgaddtomacro\gplbacktext{%
      \csname LTb\endcsname%
      \put(814,704){\makebox(0,0)[r]{\strut{}$0$}}%
      \put(814,1151){\makebox(0,0)[r]{\strut{}$0.2$}}%
      \put(814,1598){\makebox(0,0)[r]{\strut{}$0.4$}}%
      \put(814,2045){\makebox(0,0)[r]{\strut{}$0.6$}}%
      \put(814,2492){\makebox(0,0)[r]{\strut{}$0.8$}}%
      \put(814,2939){\makebox(0,0)[r]{\strut{}$1$}}%
      \put(1330,484){\makebox(0,0){\strut{}$16$}}%
      \put(2254,484){\makebox(0,0){\strut{}$32$}}%
      \put(3178,484){\makebox(0,0){\strut{}$64$}}%
      \put(4102,484){\makebox(0,0){\strut{}$128$}}%
    }%
    \gplgaddtomacro\gplfronttext{%
      \csname LTb\endcsname%
      \put(176,1821){\rotatebox{-270}{\makebox(0,0){\strut{}efficiency}}}%
      \put(2794,154){\makebox(0,0){\strut{}cores}}%
      \put(2794,3269){\makebox(0,0){\strut{}Efficiency vs number of cores (\mtop)}}%
      \csname LTb\endcsname%
      \put(2433,1317){\makebox(0,0)[r]{\strut{}\pmtwotwo}}%
      \csname LTb\endcsname%
      \put(2433,1097){\makebox(0,0)[r]{\strut{}\catfourtwo}}%
      \csname LTb\endcsname%
      \put(2433,877){\makebox(0,0)[r]{\strut{}\keightnine}}%
    }%
    \gplbacktext
    \put(0,0){\includegraphics{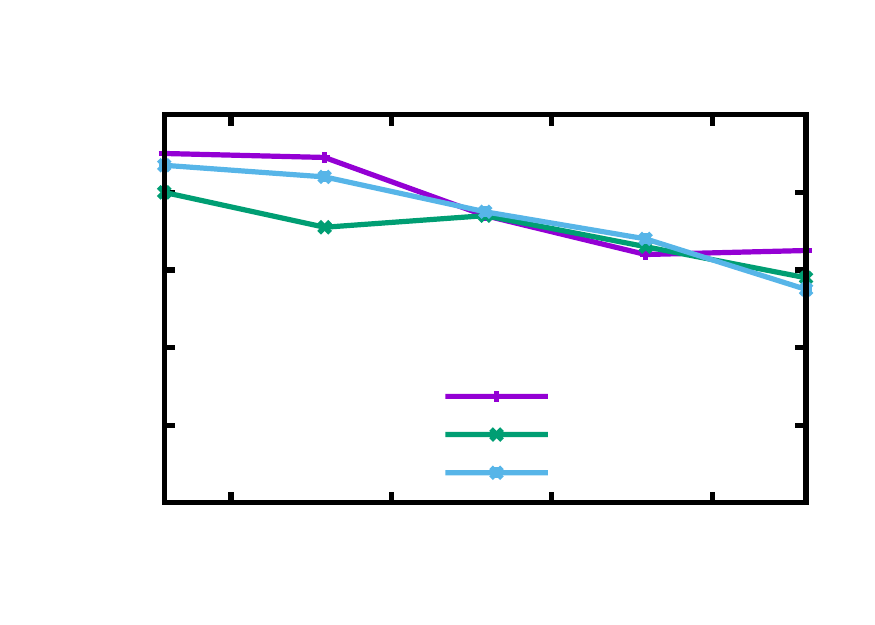}}%
    \gplfronttext
  \end{picture}%
\endgroup

%% file: plots/efficiency-mtree.tex
\begingroup
  \makeatletter
  \providecommand\color[2][]{%
    \GenericError{(gnuplot) \space\space\space\@spaces}{%
      Package color not loaded in conjunction with
      terminal option `colourtext'%
    }{See the gnuplot documentation for explanation.%
    }{Either use 'blacktext' in gnuplot or load the package
      color.sty in LaTeX.}%
    \renewcommand\color[2][]{}%
  }%
  \providecommand\includegraphics[2][]{%
    \GenericError{(gnuplot) \space\space\space\@spaces}{%
      Package graphicx or graphics not loaded%
    }{See the gnuplot documentation for explanation.%
    }{The gnuplot epslatex terminal needs graphicx.sty or graphics.sty.}%
    \renewcommand\includegraphics[2][]{}%
  }%
  \providecommand\rotatebox[2]{#2}%
  \@ifundefined{ifGPcolor}{%
    \newif\ifGPcolor
    \GPcolortrue
  }{}%
  \@ifundefined{ifGPblacktext}{%
    \newif\ifGPblacktext
    \GPblacktexttrue
  }{}%
  \let\gplgaddtomacro\g@addto@macro
  \gdef\gplbacktext{}%
  \gdef\gplfronttext{}%
  \makeatother
  \ifGPblacktext
    \def\colorrgb#1{}%
    \def\colorgray#1{}%
  \else
    \ifGPcolor
      \def\colorrgb#1{\color[rgb]{#1}}%
      \def\colorgray#1{\color[gray]{#1}}%
      \expandafter\def\csname LTw\endcsname{\color{white}}%
      \expandafter\def\csname LTb\endcsname{\color{black}}%
      \expandafter\def\csname LTa\endcsname{\color{black}}%
      \expandafter\def\csname LT0\endcsname{\color[rgb]{1,0,0}}%
      \expandafter\def\csname LT1\endcsname{\color[rgb]{0,1,0}}%
      \expandafter\def\csname LT2\endcsname{\color[rgb]{0,0,1}}%
      \expandafter\def\csname LT3\endcsname{\color[rgb]{1,0,1}}%
      \expandafter\def\csname LT4\endcsname{\color[rgb]{0,1,1}}%
      \expandafter\def\csname LT5\endcsname{\color[rgb]{1,1,0}}%
      \expandafter\def\csname LT6\endcsname{\color[rgb]{0,0,0}}%
      \expandafter\def\csname LT7\endcsname{\color[rgb]{1,0.3,0}}%
      \expandafter\def\csname LT8\endcsname{\color[rgb]{0.5,0.5,0.5}}%
    \else
      \def\colorrgb#1{\color{black}}%
      \def\colorgray#1{\color[gray]{#1}}%
      \expandafter\def\csname LTw\endcsname{\color{white}}%
      \expandafter\def\csname LTb\endcsname{\color{black}}%
      \expandafter\def\csname LTa\endcsname{\color{black}}%
      \expandafter\def\csname LT0\endcsname{\color{black}}%
      \expandafter\def\csname LT1\endcsname{\color{black}}%
      \expandafter\def\csname LT2\endcsname{\color{black}}%
      \expandafter\def\csname LT3\endcsname{\color{black}}%
      \expandafter\def\csname LT4\endcsname{\color{black}}%
      \expandafter\def\csname LT5\endcsname{\color{black}}%
      \expandafter\def\csname LT6\endcsname{\color{black}}%
      \expandafter\def\csname LT7\endcsname{\color{black}}%
      \expandafter\def\csname LT8\endcsname{\color{black}}%
    \fi
  \fi
    \setlength{\unitlength}{0.0500bp}%
    \ifx\gptboxheight\undefined%
      \newlength{\gptboxheight}%
      \newlength{\gptboxwidth}%
      \newsavebox{\gptboxtext}%
    \fi%
    \setlength{\fboxrule}{0.5pt}%
    \setlength{\fboxsep}{1pt}%
\begin{picture}(5040.00,3600.00)%
    \gplgaddtomacro\gplbacktext{%
      \csname LTb\endcsname%
      \put(814,704){\makebox(0,0)[r]{\strut{}$0$}}%
      \put(814,1151){\makebox(0,0)[r]{\strut{}$0.2$}}%
      \put(814,1598){\makebox(0,0)[r]{\strut{}$0.4$}}%
      \put(814,2045){\makebox(0,0)[r]{\strut{}$0.6$}}%
      \put(814,2492){\makebox(0,0)[r]{\strut{}$0.8$}}%
      \put(814,2939){\makebox(0,0)[r]{\strut{}$1$}}%
      \put(1330,484){\makebox(0,0){\strut{}$16$}}%
      \put(2254,484){\makebox(0,0){\strut{}$32$}}%
      \put(3178,484){\makebox(0,0){\strut{}$64$}}%
      \put(4102,484){\makebox(0,0){\strut{}$128$}}%
    }%
    \gplgaddtomacro\gplfronttext{%
      \csname LTb\endcsname%
      \put(176,1821){\rotatebox{-270}{\makebox(0,0){\strut{}efficiency}}}%
      \put(2794,154){\makebox(0,0){\strut{}cores}}%
      \put(2794,3269){\makebox(0,0){\strut{}Efficiency vs number of cores (\mtree)}}%
      \csname LTb\endcsname%
      \put(2433,1757){\makebox(0,0)[r]{\strut{}\eightcage}}%
      \csname LTb\endcsname%
      \put(2433,1537){\makebox(0,0)[r]{\strut{}\pfivecfive}}%
      \csname LTb\endcsname%
      \put(2433,1317){\makebox(0,0)[r]{\strut{}\cfivecfive}}%
      \csname LTb\endcsname%
      \put(2433,1097){\makebox(0,0)[r]{\strut{}\ksevenseven}}%
      \csname LTb\endcsname%
      \put(2433,877){\makebox(0,0)[r]{\strut{}\konetwo}}%
    }%
    \gplbacktext
    \put(0,0){\includegraphics{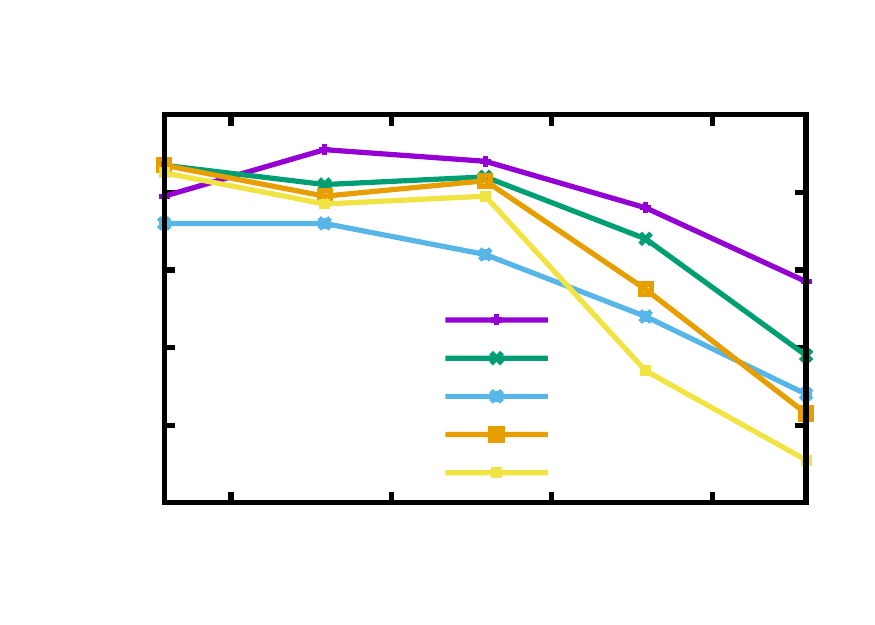}}%
    \gplfronttext
  \end{picture}%
\endgroup